\documentclass[12pt]{article}
\pdfoutput=1

\usepackage{amsmath,amssymb,amsthm,graphicx,amscd,multirow,xspace}
\usepackage[colorlinks=true,urlcolor=blue,anchorcolor=blue,citecolor=blue,filecolor=blue,linkcolor=blue,menucolor=blue,pagecolor=blue]{hyperref}
\usepackage[compress,numbers]{natbib}
\usepackage{subfigure, placeins}
\usepackage{pdflscape}
\usepackage[all]{xy}
\usepackage{color}
\usepackage[usenames,dvipsnames,svgnames,table]{xcolor}
\usepackage{multirow}
\usepackage{soul,slashed}
\usepackage{bm}
\usepackage{dsfont}
\usepackage{mathrsfs}
\usepackage{bbold}
\usepackage{floatrow}
\newfloatcommand{capbtabbox}{table}[][\FBwidth]
\usepackage{arydshln}

\long\def\symbolfootnote[#1]#2{\begingroup%
\def\thefootnote{\fnsymbol{footnote}}\footnote[#1]{#2}\endgroup}

\newcommand{\newc}{\newcommand}
\newc{\gsim}{\lower.7ex\hbox{$\;\stackrel{\textstyle>}{\sim}\;$}}
\newc{\lsim}{\lower.7ex\hbox{$\;\stackrel{\textstyle<}{\sim}\;$}}
\newc{\gev}{\,{\rm GeV}}
\newc{\mev}{\,{\rm MeV}}
\newc{\ev}{\,{\rm eV}}
\newc{\kev}{\,{\rm keV}}
\newc{\tev}{\,{\rm TeV}}

\newc{\mz}{M_Z}
\newc{\mpl}{M_*}
\newc{\mw}{m_{\rm weak}}
\newc{\nr}[1]{N^c_R{}_{#1}}

\newcommand{\One}{\mathbb{1}}
\def\be{\begin{equation}}
\def\ee{\end{equation}}

\def\CN{{\mathcal{N}}}

\def\IZ{{\mathbb{Z}}}

\def\IR{{\mathbb{R}}}

\def\One{{\mathbb{1}}}

\theoremstyle{definition}

\def\beq{\begin{equation}}
\def\eeq{\end{equation}}
\def\bea{\begin{eqnarray}}
\def\eea{\end{eqnarray}}
\def\bitem{\begin{itemize}}
\def\eitem{\end{itemize}}
\newc{\ie}{{\it i.e.}}          \newc{\etal}{{\it et al.}}
\newc{\eg}{{\it e.g.}}          \newc{\etc}{{\it etc.}}
\newc{\cf}{{\it c.f.}}

\newcommand{\Z}{\mathbb{Z}}

 \numberwithin{equation}{section}

\newcommand\fverb{\setbox\fverbbox=\hbox\bgroup\verb}
\newcommand\fverbdo{\egroup\medskip\noindent%
            \fbox{\unhbox\fverbbox}\ }
\newcommand\fverbit{\egroup\item[\fbox{\unhbox\fverbbox}]}
\newbox\fverbbox

\usepackage{authblk}

\author[1]{Nathaniel Craig\thanks{ncraig@physics.ucsb.edu}}
\author[2,3]{Simon Knapen\thanks{smknapen@lbl.gov}}
\author[4]{Pietro Longhi\thanks{pietro.longhi@physics.uu.se}}
\author[5]{Matthew Strassler\thanks{strassler@physics.harvard.edu}}

\affil[1]{\small{Department of Physics, University of California, Santa Barbara, CA 93106}}
\affil[2]{Berkeley Center for Theoretical Physics,
University of California, Berkeley, CA 94720
}
\affil[3]{ Theoretical Physics Group, Lawrence Berkeley National Laboratory, Berkeley, CA 94720
}
\affil[4]{Department of Physics and Astronomy, Uppsala University, Box 516, SE-75120 Sweden}
  
  \affil[5]{Department of Physics, Harvard University, Cambridge, MA 02138}

\title{The Vector-like Twin Higgs}

\begin{document}

\maketitle
\begin{abstract}
  We present a version of the twin Higgs mechanism with vector-like top partners.  In this setup all gauge anomalies automatically cancel, even without twin leptons. The matter content of the most minimal twin sector is therefore just two twin tops and one twin bottom. The LHC phenomenology, illustrated with two example models, is dominated by twin glueball decays, possibly in association with 
  Higgs bosons. We further construct an explicit four-dimensional UV completion and discuss a variety of UV completions relevant for both vector-like and fraternal twin Higgs models.

\end{abstract}

\clearpage
\tableofcontents

\section{Introduction}

The non-observation of new physics at Run 1 of the LHC poses a sharp challenge to conventional approaches to the hierarchy problem. The challenge is particularly acute due to stringent limits on fermionic and scalar top partners, which are expected to be light in symmetry-based solutions to the hierarchy problem such as supersymmetry or compositeness. Bounds on these top partners rely not on their intrinsic couplings to the Higgs, but rather their QCD production modes, which arise when the protective symmetries commute with Standard Model gauge interactions. However, the situation can be radically altered when approximate or exact discrete symmetries play a role in protecting the weak scale \cite{Chacko:2005pe,Chacko:2005un, Craig:2014roa,Craig:2014aea}. In this case the lightest states protecting the Higgs can be partially or entirely neutral under the Standard Model, circumventing existing searches while giving rise to entirely new signs of naturalness.

The twin Higgs \cite{Chacko:2005pe,Chacko:2005un} is the archetypal example of a theory where discrete symmetries give rise to partner particles neutral under the Standard Model. Here the weak scale is protected by a $\mathbb{Z}_2$ symmetry relating the Standard Model to a mirror copy; the discrete symmetry may be exact or a residual of more complicated dynamics \cite{Craig:2014roa,Craig:2014aea, Geller:2014kta, Barbieri:2015lqa, Low:2015nqa}. In the twin Higgs and its relatives, both the Standard Model and the twin sector are chiral, with fermions obtaining mass only after spontaneous symmetry breaking. If the $\mathbb{Z}_2$ symmetry is exact, this fixes the mass spectrum of the twin sector uniquely in terms of the symmetry breaking scale $f$. Even if the $\mathbb{Z}_2$ is not exact, naturalness considerations fix the mass of the twin top quark in terms of $f$, while the masses of other twin fermions should be significantly lighter. \cite{Craig:2015pha}. 

In this respect the twin Higgs is qualitatively different from conventional theories involving supersymmetry or continuous global symmetries, in which the masses of nearly all partner particles may be lifted by additional terms without spoiling the cancellation mechanism. This allows states irrelevant for naturalness to be kinematically decoupled, as in the paradigm of natural SUSY \cite{Dimopoulos:1995mi,Cohen:1996vb}.  As we will show, the cancellation mechanism of the twin Higgs is not spoiled by the presence of vector-like masses for fermions in the twin sector, as these mass terms represent only a soft breaking of the twin symmetry. This raises the prospect that partner fermions in the twin sector may acquire vector-like masses, significantly altering the phenomenology of (and constraints on) twin theories.  Moreover due to the vector-like nature of the twin fermions, twin leptons are no longer needed to cancel the gauge anomalies in the twin sector \cite{Craig:2014roa}. Any tension with cosmology is therefore trivially removed.

The collider phenomenology of this class of models has a few important new features.  While it resembles the `fraternal twin Higgs' \cite{Craig:2015pha} (in that the 125 GeV Higgs may decay to twin hadrons with measurable branching fractions, and the decays of the twin hadrons to Standard Model particles may occur promptly or with displaced vertices), the role of the radial mode of the Higgs potential can be more dramatic than in the fraternal case.  Not only are twin hadrons more often produced in radial mode decays, because of the absence of light twin leptons, but also flavor-changing currents in the twin sector can lead to a new effect: {\it emission of on- or off-shell Higgs bosons.} 
Searches for very rare events with one or more Higgs bosons 
or low-mass non-resonant $b\bar b$ or $\tau^+\tau^-$ pairs, 
generally accompanied by twin hadron decays and/or missing energy, are thus motivated by these models.    Other interesting details in the twin hadron phenomenology can arise, though the search strategies just mentioned -- and those appropriate for the fraternal twin Higgs -- seem sufficient to cover them.

Although a vector-like spectrum of twin fermions appears compatible with the cancellation mechanism of the twin Higgs, it raises a puzzling question: What is the fundamental symmetry? A vector-like twin sector entails additional matter representations not related to the Standard Model by an obvious $\mathbb{Z}_2$ exchange symmetry. In this case it is no longer obvious that the Standard Model and twin sectors share the same cutoff $\Lambda$. The vector-like spectrum also necessarily entails unequal contributions to the running of twin sector gauge couplings, so that the cancellation mechanism will be spoiled at two loops. This requires that the vector-like twin Higgs resolve into (at least) a $\mathbb{Z}_2$-symmetric UV completion in the range of 5-10 TeV. The emergence of approximate IR $\mathbb{Z}_2$ symmetries from more symmetric UV physics is a natural ingredient of orbifold Higgs models \cite{Craig:2014roa,Craig:2014aea}. As we will see, orbifold Higgs models inspire suitable UV completions of the vector-like twin Higgs in four or more dimensions. As a by-product, we provide a straightforward way to UV complete the spectrum of the fraternal twin Higgs in \cite{Craig:2015pha}. Note also that a vector-like mass spectrum has a natural realization in the Holographic Twin Higgs \cite{Geller:2014kta}, where spontaneous breaking of a bulk symmetry leads to modest masses for twin sector fermions.

This paper is organized as follows: In Section \ref{sec:vectortwin} we introduce a toy vector-like extension of the twin Higgs and show that it protects the weak scale in much the same way as the chiral twin Higgs. 
In Section \ref{sec:models} we present a minimal example of a complete vector-like twin model, as well as a second, non-minimal model. The former is the vector-like analogue of the fraternal twin Higgs, and provides an equally minimal realization of the twin mechanism.
The phenomenological implications of both models are discussed in Section \ref{sec:pheno}. We address the question of fundamental symmetries in Section \ref{sec:uv}, providing both explicit 4D models inspired by dimensional deconstruction and their corresponding orbifold constructions. We conclude in Section \ref{sec:conc}. In Appendix \ref{appHypercharge} we include a new way to deal with hypercharge in orbifold Higgs models.

\section{The Vector-like Twin Higgs} \label{sec:vectortwin} 

 In this section we review the twin Higgs and introduce our generalization of it, treating the top quark and Higgs sector as a module or toy model.  We will explore more complete models in section \ref{sec:models}.

In the original twin Higgs, the Standard Model is extended to include a complete mirror copy whose couplings are related to their Standard Model counterparts by a $\mathbb{Z}_2$ exchange symmetry. In a linear sigma model realization of the twin Higgs, the interactions of the Higgs and the top sector take the form 
 \begin{equation}
\begin{split}\label{twinlag}
-\mathcal{L}\supset &-m^2 \Big[|H|^2+|H'|^2\Big]+\lambda \Big[|H|^2+|H'|^2\Big]^2+\delta \Big[|H|^4+|H'|^4\Big]\\
& +y_t\,  H\, q\,u+ y_t\,  H' q'u' + {\rm h.c.}
\end{split}
\end{equation}
with $\lambda,\delta>0$ and where $H$ and $q, u$ are the Higgs doublet and the third generation up-type quarks charged under the Standard Model gauge interactions. Similarly, the primed fields denote the twin sector analogues of these fields, charged under the twin sector gauge group.

The first two terms in (\ref{twinlag}) respect an $SU(4)$ global symmetry, while the remaining dimensionless terms exhibit the $\mathbb{Z}_2$ symmetry exchanging the primed and unprimed fields. This $\mathbb{Z}_2$ leads to radiative corrections to the quadratic action that respect the $SU(4)$ symmetry. Indeed, a simple one-loop computation with $\mathbb{Z}_2$-symmetric cutoff $\Lambda$ gives a correction to the Higgs potential of the form
\begin{eqnarray} \label{eq:cws3}
-\mathcal{L}^{(1)} \supset \frac{\Lambda^2}{16 \pi^2} \Big( - 6 y_t^2 + \frac{9}{4} g_2^2 +10 \lambda + 6  \delta \Big)  \Big(  |H|^2 +|H'|^2 \Big)\,.
\end{eqnarray}
 The effective potential possesses the customary $SU(4)$ symmetric form, so that a goldstone of spontaneous $SU(4)$ breaking may remain protected against one-loop sensitivity to the cutoff. 

 When $H$ and $H'$ acquire vacuum expectation values, they spontaneously break the accidental $SU(4)$ symmetry, giving rise to a pseudo-goldstone scalar $h$ identified with the Standard Model-like Higgs. This pNGB is parametrically lighter than the radial mode associated with the breaking of the  accidental $SU(4)$, provided that $\delta \ll \lambda$. 
 
 Note that the potential (\ref{daughertree}) leads to vacuum expectation values $v = v' = f/\sqrt{2}$. Unequal vevs -- and a pNGB Higgs aligned mostly with the SM vev -- can be obtained by introducing a soft $\mathbb{Z}_2$-breaking mass parameter $\delta m$, such that $v \ll v' \sim f$ occurs with a $\mathcal{O}(v^2/ 2 f^2)$ tuning of parameters. The current status of precision Higgs coupling measurements requires $v/f\lesssim 1/3$, see for instance \cite{Burdman:2014zta}.

The sense in which twin top quarks serve as top partners is clear if we integrate out the heavy radial mode of accidental $SU(4)$ breaking. This can be most easily done by using the identity
\begin{equation}
|H|^2+|H'|^2=f^2/2
\end{equation}
to solve for $H'$. In the unitary gauge, this then gives rise to couplings between the pNGB Higgs and fermions of the form
\begin{equation}\label{eq:lagPNBG}
-\mathcal{L}\supset  \frac{1}{\sqrt{2}}\,y_t\,  (v+h)\,  q\,{u} + \frac{1}{\sqrt{2}}\,y_t\,\left(f-\frac{1}{2 f} (v+h)^2 \right) q'u' + \dots
\end{equation}
where $h$ is the physical Higgs boson and the trailing dots indicate $v^3/f^3$ suppressed corrections. 
These are precisely the couplings required to cancel quadratic sensitivity of the pNGB Higgs to higher scales, provided the cutoff is $\mathbb{Z}_2$-symmetric.

The vector-like twin Higgs entails the extension of this twin sector to include fermions transforming in vector-like representations of the twin gauge group. The vector-like extension of (\ref{twinlag}) is then
\begin{equation}
\begin{split}\label{daughertree}
-\mathcal{L}\supset &-m^2 \Big[|H|^2+|H'|^2\Big]+\lambda \Big[|H|^2+|H'|^2\Big]^2+\delta \Big[|H|^4+|H'|^4\Big]\\
& +y_t\,  H\, q\,u+ y_t\,  H' q'u' + M_Q\,q'  \bar q' +M_U\, u'  \bar u' + {\rm h.c.}
\end{split}
\end{equation}
where we have introduced additional fields $\bar q'$ and $\bar u'$ that are vector-like partners of the twin tops. The generalization to multiple generations, as well as the down-type quark and lepton sectors is again straightforward, and is discussed in detail in the next section. 

Although the additional fermions and vector-like mass terms $M_{Q,U}$ break the $\mathbb{Z}_2$ symmetry, they do so softly and thus do not reintroduce a quadratic sensitivity to the cut-off. Quadratically divergent contributions to the Higgs potential are still proportional to an $SU(4)$ invariant as in (\ref{eq:cws3}), assuming equal cutoffs for the two sectors.

There are several points worth emphasizing about this cancellation. First, note that the apparent symmetries of the vector-like twin Higgs also allow additional operators which we have not yet discussed. There are possible Yukawa couplings of the form
\begin{equation}\label{eq:wrongHiggscoupling}
\mathcal{L} \supset \tilde y_t H^{\prime \dag} \bar q' \bar u' + {\rm h.c.}
\end{equation}
These couplings, if large, provide additional radiative corrections to the potential for $H'$ that would spoil the twin cancellation mechanism. While it is technically natural to have $\tilde y_t\ll 1$, there are also several ways of explicitly suppressing this coupling: For instance, in a supersymmetric UV completion, (\ref{eq:wrongHiggscoupling}) is forbidden by holomorphy. Alternatively, in a (deconstructed) extra dimension there could be some geographical separation between $H'$ and $\bar q',\bar u'$, which would also suppress this Yukawa coupling.  Finally (\ref{eq:wrongHiggscoupling}) can be forbidden by a PQ symmetry, which is softly broken by $M_Q$ and $M_U$. In section \ref{sec:uv} we will present an explicit UV completion which implements the first two ideas.  Another set of operators, of the form
\begin{equation}\label{eq:dangerousone}
\mathcal{L} \supset  c \frac{ M_Q}{\Lambda^2} HH^\dagger \bar q' q'+\mathrm{etc} \ ,
\end{equation}
can lead to a one loop contribution to the Higgs mass of the form
\begin{equation}
\delta m_h^2 \sim \frac{c}{16\pi^2} M_Q^2.
\end{equation}
In perturbative UV completions one generally expects $c\sim 1$ or $c\ll 1$, which renders \eqref{eq:dangerousone} subleading with respect to a set of logarithmic corrections which we will discuss shortly.  (In the supersymmetric UV completions we provide in section \ref{sec:uv}, $c\ll1$.)  In strongly coupled UV completions, it could happen that $c\sim 16\pi^2$, which would require  $M_Q\lesssim m_h$.  But $c$ can be suppressed below the NDA estimate by a selection rule, or by the strong dynamics itself, as for instance through a geographical separation between $H'$ and $\bar q'$ in a warped extra dimension.

Second, the additional vector-like fermions change the running of twin sector gauge couplings, which in turn cause twin-sector Yukawa couplings to deviate from their Standard Model counterparts. The most important effect is in the running of the QCD and $\mathrm{QCD}'$ gauge couplings, which in the presence of three full generations of vector-like twin quarks take the form
\begin{equation}
\begin{split}
\beta_{g_3} &= -7 \frac{g_3^3}{16 \pi^2} + \mathcal{O}(g_3^5) \\
\hspace{1cm} \beta_{g_3'} &= -3 \frac{g_3^{\prime 3}}{16 \pi^2} + \mathcal{O}(g_3^{\prime 5}) \,.
\end{split}
\end{equation}
The mismatch in the QCD beta-functions also induces a tiny two-loop splitting between the SM and twin top Yuwaka couplings at the weak scale.  But cancellation of quadratically divergent contributions to the Higgs mass is computed at the scale $\Lambda$, so that the different running of the strong gauge and Yukawa couplings causes no problem as long as the physics of the UV completion at $\Lambda$ is $\mathbb{Z}_2$ symmetric. This implies, at the very least, that the model must be UV completed into a manifestly $\Z_2$ symmetric setup at a relatively low scale.

\begin{figure}
\includegraphics[width=\textwidth]{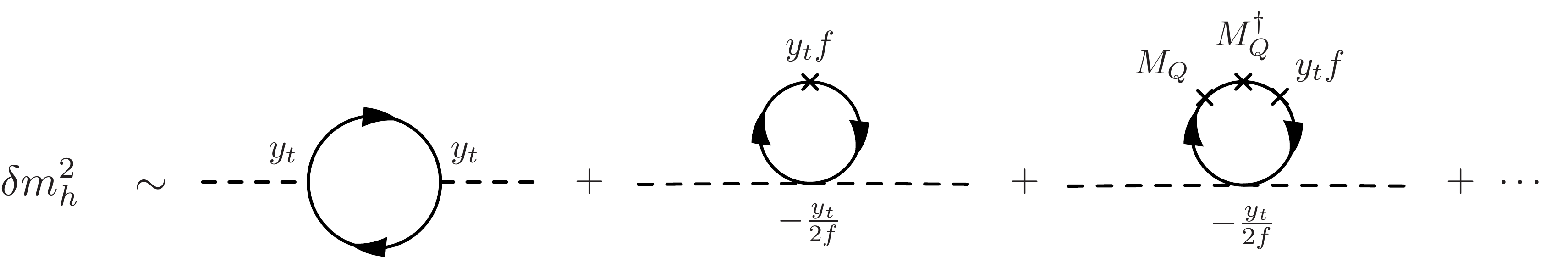}
\caption{Diagrams correcting the pseudo-goldstone mode.\label{softdiagrams}}
\end{figure}

Although cutoff sensitivity is still eliminated at one loop, the vector-like masses will result in log-divergent threshold corrections to the Higgs mass that must be accounted for in the tuning measure. To see these features explicitly, it is useful to  again work in the low-energy effective theory obtained by integrating out the radial mode of $SU(4)$ breaking in the twin Higgs potential. This now gives
\begin{equation}\label{eq:vectortwinaction}
-\mathcal{L}\supset\frac{y_t}{\sqrt{2}}\,  (h+v)\,  q\,{u} + \frac{y_t}{\sqrt{2}}\left(f-\frac{1}{2 f} (h+v)^2 \right) q'u' +M_Q \,q'  \bar q' + M_U\, u'  \bar u' + \dots
\end{equation}
The only difference with the conventional twin Higgs is the presence of the vectorlike mass terms. From a diagrammatic point of view, it is now easy to see that the leading quadratic divergence exactly cancels as it does in the regular twin Higgs. Moreover any diagrams with additional $M_Q$ and $M_U$ mass terms must involve at least two such insertions, which is sufficient to soften the diagram enough to make it logarithmically divergent (see Fig.~\ref{softdiagrams}). Concretely, this implies log-divergent  contributions to the Higgs mass parameter $m_{ h}^2$ of the form 
\begin{equation}
\delta m_{h}^2 \sim \frac{3 y_{t}^2}{4 \pi^2} \left( M^2_{Q}\log\left[\frac{ M^2_{Q}}{\Lambda^2}\right]+ M_{U}^2 \log\left[\frac{ M^2_{U}}{\Lambda^2}\right] \right) 
\end{equation}
Unsurprisingly, this constrains the vector masses by the requirement that the threshold corrections to $m_{h}$ not be too large, meaning $M_{Q}, M_{U}\lesssim 450$ GeV. 
\footnote{
{
One may wonder if this source of $\IZ_2$ breaking could naturally generate the $v\ll f$ hierarchy. 
This is not the case, as it comes with the wrong sign. An additional source of soft $\mathbb{Z}_2$ breaking therefore remains necessary.
}
}

Although the impact of a vector-like twin sector on the twin cancellation mechanism is relatively minor, the effects on phenomenology are much more radical. First and foremost, the vector-like twin top sector, as presented in this section, is anomaly free by itself and therefore constitutes the simplest possible self-consistent vector-like twin sector. In this sense it is the vector-like analogue of the fraternal twin Higgs \cite{Craig:2015pha}, \emph{but without the need for a twin tau and twin tau neutrino}. In terms of minimality, this places lepton-free vector-like twin Higgs models on comparable footing with the fraternal twin Higgs. Secondly, in the presence of multiple generations of twin quarks, the $M_{Q,U}$ are promoted to matrices in flavor space. The twin flavor textures of these vector-like mass terms are not necessarily aligned with that of the Yukawa, such that one generically expects large flavor changing interactions in the twin sector, which may lead to interesting collider signatures.

\section{Example Models\label{sec:models}}
As argued in \cite{Craig:2015pha}, naturalness of the Higgs potential allows for a substantial amount of freedom in the choice of the field content and couplings of the twin sector. 
In the vector-like twin Higgs this freedom is even greater, and results in a large class of models featuring rich and diverse phenomenology. 
Aside from the Higgs sector introduced in the previous section, all models contain a twin sector with the following components:
\begin{itemize}
\item \textbf{Gauge sector:}  A twin  $SU(2)\times SU(3)$ gauge symmetry is necessary for naturalness, although the difference between the twin gauge couplings and their Standard Model counterparts can be of the order of $\delta g_{2,3}/g_{2,3}\sim 10$\%, evaluated at the scale $\Lambda$ \cite{Craig:2015pha}. In particular this implies that the confinement scale of the twin QCD sector may vary within roughly an order of magnitude. Twin hypercharge does not significantly impact the fine tuning and may be omitted from the model.  We will leave the twin $U(1)$ ungauged in what follows, with the consequence of degenerate twin electroweak gauge bosons, which we denote with $W'$ and $Z'$. We do however assume that twin hypercharge is present as a global symmetry, and as such it imposes selection rules on the decays of the quarks. 

\item \textbf{Top sector:} In the top sector naturalness demands that we include the twin partner of the Standard Model top and that the top and twin-top Yukawa couplings differ by no more than about 1\%. We must also introduce the left-handed twin bottom, as it forms a doublet with the left-handed twin top. The key difference with the conventional twin Higgs is that these twin partners are now Dirac rather than Weyl. As argued in the previous section, to preserve naturalness the corresponding Dirac mass terms  should also not exceed $\sim 500$ GeV.

\item \textbf{Quark sector:} The remaining quarks are all optional, as they are required neither for naturalness nor anomaly cancellation. If they are present, they can have vector-like masses as heavy as  $\sim 5$ TeV, which corresponds to the cut-off of the effective theory. In this case the UV completion must provide some form of flavor alignment between the Yukawa's and the vector-like mass terms, but as we will see, this is generally not difficult to achieve.

\item \textbf{Lepton sector:} Unlike in chiral versions of the twin Higgs, twin leptons are not required for anomaly cancellation and are therefore optional as well. If present, they too can be taken heavy, and therefore easily by-pass any cosmological constraints on the number of relativistic degrees of freedom.  

\end{itemize}

The parameter space is too large for us to study  in full generality, so instead we study two well-motivated cases:
\begin{itemize}
\item \textbf{Minimal vector-like model:} We consider the most minimal twin sector required by naturalness, consisting of a single vector-like generation of twin (top) quarks. This model is therefore the vector-like analogue of the fraternal twin Higgs \cite{Craig:2015pha}, with the crucial difference that twin leptons are absent entirely. We will show that it shares many phenomenological features with the fraternal twin Higgs.

\item  \textbf{Three-generation model:} 
In this model we include the partners of all SM fermions, but we effectively decouple the twin partners of the $\mathbf{ \overline 5}$ multiplet ($d$, $\ell$), by setting their vector-like masses well above the top partner mass $y_t f$. The twin partners of the $\mathbf{ 10}$ ($q$, $u$, $e$) remain near the weak scale, a spectrum which arises naturally in the most simple UV completions (see section \ref{sec:simpleuv}). While we do allow for flavor-generic Dirac masses for the remaining quarks, we take all entries of the mass matrices $\lesssim f/\sqrt{2}$ to preserve naturalness. The right-handed twin leptons may also be  in the few-hundred GeV range, but in the absence of twin hypercharge they decouple completely from the phenomenology, and we will not discuss them further.
\end{itemize}
In the remainder of this section we will study the spectrum of these two cases, with a focus on the constraints imposed by naturalness. We reserve a detailed study of their collider signatures for  section \ref{sec:pheno}. For UV completions of both scenarios we refer to section \ref{sec:uv}.

\subsection{Minimal vector-like model\label{sec:fraternal}}

\begin{figure}[!t]
\includegraphics[width=0.47\textwidth]{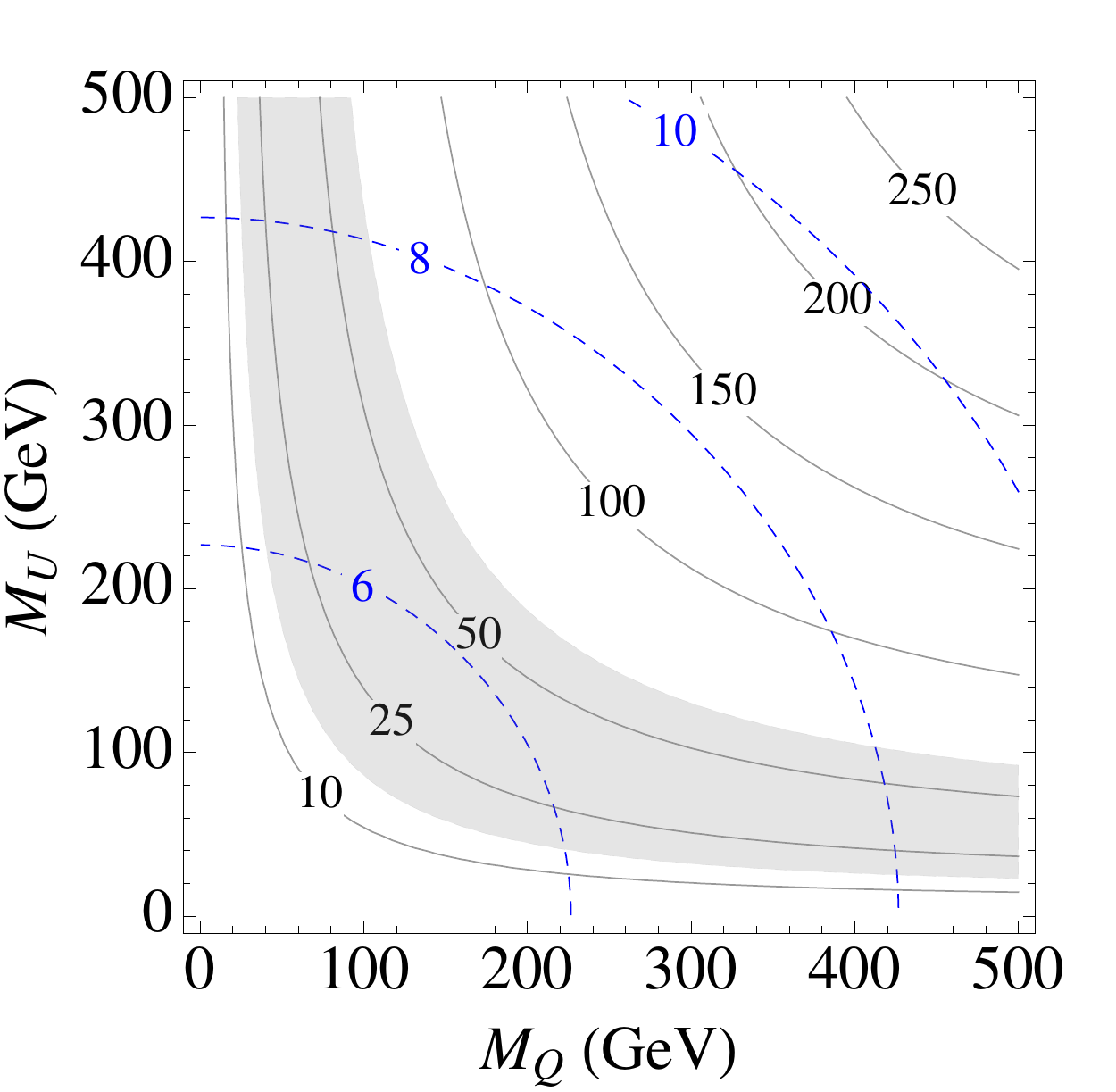}\hfill
\includegraphics[width=0.47\textwidth]{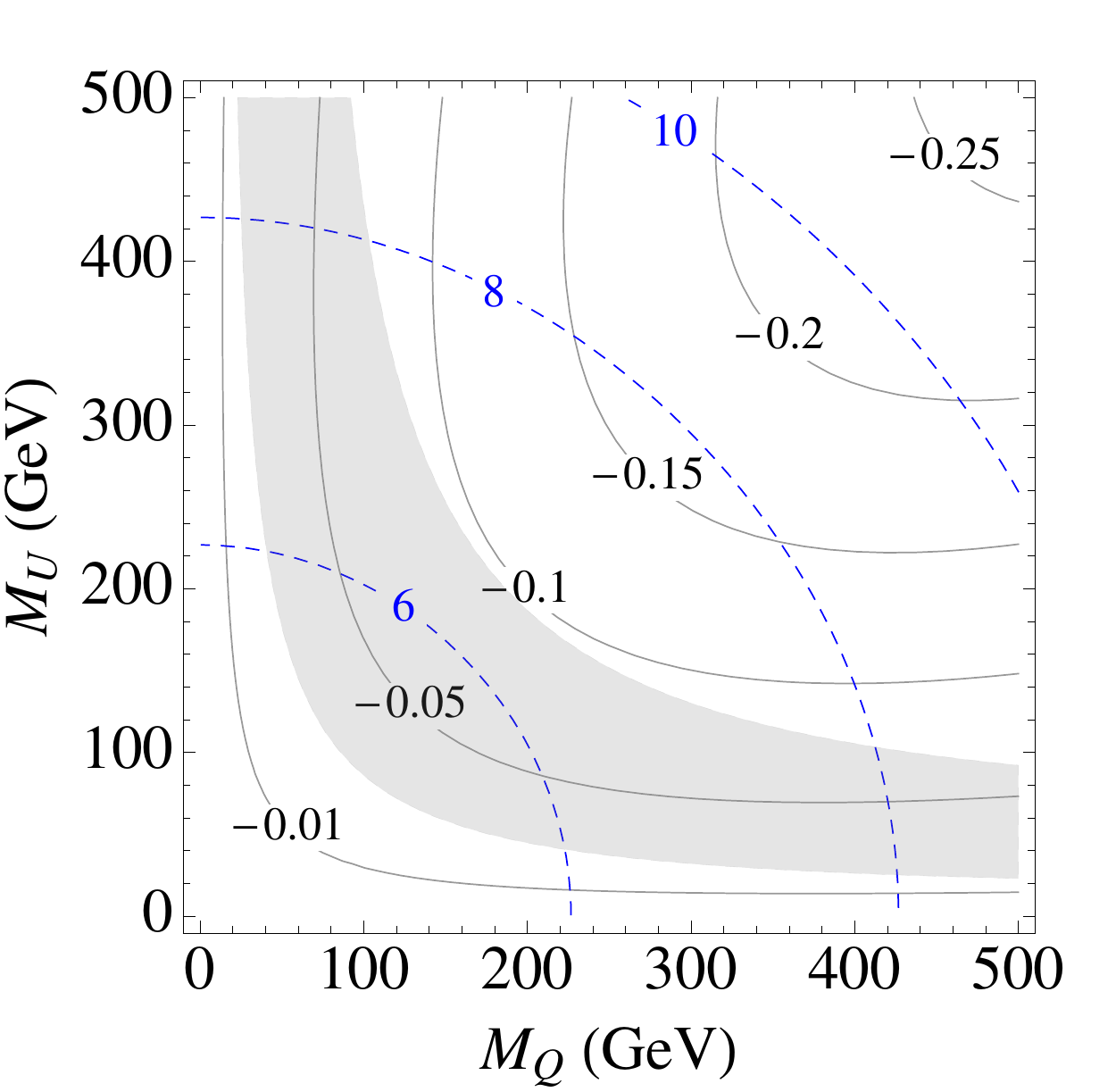}
\caption{\label{fig:mt1} Plots of the $m_{t_1}$ (left) and ${\scriptstyle  {\mathcal Y}}_{11}$ (right) of the lightest twin top as a function of $M_Q$ and $M_U$ with $f=750$ GeV (black lines). Dashed blue lines lines indicate approximate fine-tuning measure $\Delta$ as a result of the threshold correction in (\ref{thresholdcor}) for $\Lambda=5$ TeV. The gray shading indicates the perturbative estimate of the region excluded by $h\rightarrow t'_1\bar t'_1$ decays, as explained in section \ref{sec:pheno}. This can however have large non-perturbative corrections; see appendix B of \cite{Craig:2015pha}. }
\end{figure}

In terms of Weyl spinors --- we will use Weyl notation for spinors throughout --- the fermion content of the twin sector is just given by 
\begin{equation}
\begin{array}{c|cccc}
&q'&\bar q'&u'&\bar u' \\\hline
SU(3)' &\square&\overline\square&\overline\square&\square\\
SU(2)'&\overline\square&\square&1&1\\
\end{array}
\end{equation}
 The Lagrangian is the one in (\ref{eq:vectortwinaction}).  As argued in  section \ref{sec:vectortwin}, the vector-like mass terms are constrained by naturalness to reside in the range $0< M_Q, M_U\lesssim y_t f/\sqrt{2} \sim (f/v)\times 170$ GeV. 
The spectrum then contains two top-like states and one bottom-like state, which we will denote with $t'_{1,2}$ and $b'_1$ respectively. The mass of the $b'_1$ state is just $m_{b_1}=M_Q$. From \eqref{eq:vectortwinaction}, the mass matrix of the top sector is given by 
\begin{equation}
-\mathcal{L}\supset\left(\!\!\begin{array}{c}\bar q'_u\\ u'\end{array}\!\!\right)^T   \left(\!\!\begin{array}{cc} M_Q&0\\\frac{y_t f}{\sqrt{2} } &M_U\end{array}\!\!\right)    \left(\!\!\begin{array}{c} q'_u\\ \bar u'\end{array}\!\!\right)
\end{equation}
where $q'_u$ ($\bar q'_u$) indicates the up component of the doublet $q'$ ($\bar q'$). We neglected the $v^2/f^2$ suppressed contribution to the lower left entry. Since $ y_t f/\sqrt{2} \gtrsim M_Q, \;M_U$, this system contains a (mini) seesaw. This implies the ordering $m_{t_2}>m_{b_1}>m_{t_1}$. The tops are moreover strongly mixed, with masses 
\begin{align}
m_{t_1}^2&=\frac{1}{2}\left(\;M_Q^2+M_U^2+\frac{1}{2}y_t^2 f^2 - \sqrt{\Big(M_Q^2+M_U^2+\frac{1}{2}y_t^2f^2\Big)^2-4M_Q^2 M_U^2}\;\right)\\
&\approx 2\frac{M_Q^2 M_U^2}{y_t^2 f^2}\\
m_{t_2}^2&=\frac{1}{2}\left(\;M_Q^2+M_U^2+\frac{1}{2}y_t^2 f^2 + \sqrt{\Big(M_Q^2+M_U^2+\frac{1}{2}y_t^2f^2\Big)^2-4M_Q^2 M_U^2}\;\right)\\
&\approx\frac{1}{2} y_t^2 f^2+M_Q^2 +M_U^2
\end{align}
where the expansion is for small $M_Q/f\sim M_U/f$.   For $f/v=3$, this implies that the heavier twin top has a mass between 500 and 600 GeV, while the lighter has a mass which can range between 10 and 200 GeV, as shown in the left-hand panel of Figure \ref{fig:mt1}. From (\ref{eq:lagPNBG}), the mass eigenstates couple to the SM Higgs as follows %
\begin{equation}
-\mathcal{L}\supset -\frac{1}{\sqrt{2}}\left(\frac{1}{2f} h^2 +\frac{v}{f} h\right)\bigg({\scriptstyle  {\mathcal Y}}_{11}\overline t_1' t_1'+{\scriptstyle  {\mathcal Y}}_{22}\overline t_2' t_2'+{\scriptstyle  {\mathcal Y}}_{12}\overline t_1' t_2'+{\scriptstyle  {\mathcal Y}}_{21}\overline t_2' t_1'\bigg)
\end{equation}
with
\begin{align}
{\scriptstyle  {\mathcal Y}}_{11}&
  =-\frac{y_t^2 f}{\sqrt{2}}\frac{m_{t_1}}{m_{t_2}^2-m_{t_1}^2}
\approx -2\frac{M_Q M_U}{y_t f^2}\label{eq:yt1t1}\\
{\scriptstyle  {\mathcal Y}}_{22}&
=\frac{y_t^2 f}{\sqrt{2}}\frac{m_{t_2}}{m_{t_2}^2-m_{t_1}^2}
\approx y_t\left( 1 - \frac{M_Q^2+M_U^2}{y_t^{2} f^2}\right)\\
{\scriptstyle  {\mathcal Y}}_{12}%
&\approx\sqrt{2}\frac{M_Q}{f}\left( 1 -3 \frac{M_U^2}{y_t^2 f^2}-\frac{M_Q^2}{y_t^2 f^2}\right)
\label{eq:yt1t2}\\
{\scriptstyle  {\mathcal Y}}_{21} %
&\approx-\sqrt{2}\frac{M_U}{f}\left( 1 - 3\frac{M_Q^2}{y_t^2 f^2}-\frac{M_U^2}{y_t^2 f^2}\right)\,.\label{eq:yt2t1}
\end{align}
where the approximate equalities again indicate an expansion in $M_Q/f$ and $M_U/f$. From (\ref{eq:yt1t1}) we see that (when its mass is small compared to $M_Q,M_U$) the $t_1'$ couples to the light Higgs with a coupling proportional to {\it minus} its mass   
\begin{equation}\label{eq:t1coupling}
-\mathcal{L}\supset \frac{v}{f}\frac{m_{t_1}}{f}\, h\, t'_{1}\, \bar t'_{1}\left(1 - 2\frac{M_Q^2+M_U^2}{f^2 y_t^2}+\cdots\right),
\end{equation}
as follows from the seesaw. This behavior is shown quantitatively in the right-hand panel of figure \ref{fig:mt1}.

At this point we can compute the correction to the SM Higgs mass in the minimal vector-like model, accounting for the mixing between the twin tops. The order-$\Lambda^2$ piece is
\begin{equation}
\delta m_h^2 = -\frac{3}{2\pi^2}\frac{-1}{{\sqrt{2} f}}\left({\scriptstyle  {\mathcal Y}}_{11} m_{t_1} +{\scriptstyle  {\mathcal Y}}_{22}m_{t_2}\right) \Lambda^2=%
+\frac{3}{4\pi^2}y_t^2 \Lambda^2
\end{equation}
which cancels against the contribution from the Standard Model top, %
as expected. The  logarithmically divergent correction is
\begin{align}\label{thresholdcor}
\delta m_h^2 &=%
 -\frac{3}{4\pi^2} y_t^2
 \left(m_t^2\log\left[\frac{m_t^2}{\Lambda^2}\right]-\frac{m_{t_2}^4\log\left[\frac{m_{t_2}^2}{\Lambda^2}\right]-m_{t_1}^4\log\left[\frac{m_{t_1}^2}{\Lambda^2}\right]}{m_{t_2}^2-m_{t_1}^2}\right)\\
  &=-\frac{3}{4\pi^2} y_t^2 m_t^2\log\left[\frac{m_t^2}{\Lambda^2}\right]+\frac{3}{4\pi^2} y_t^2\big(m_{t_2}^2+m_{t_1}^2\big)\log\left[\frac{m_{t_2}^2}{\Lambda^2}\right]+\mathcal{O}\left(\frac{m_{t_1}^4}{m_{t_2}^2}\right)\label{thresholdcor2}
\end{align} 
again up to $v^2/f^2$ suppressed contributions. 
The first term in (\ref{thresholdcor2}) is just the contribution from the Standard Model top, whose mass is denoted by $m_t$. 
In the limit where we turn off the vector-like masses $M_Q,M_U\to 0$, we have $m_{t_1}\rightarrow 0$ and $m_{t_2}\rightarrow \frac{1}{\sqrt{2}}y_t f$. The lightest twin top then ceases to contribute to (\ref{thresholdcor}), while the contribution of the heavier twin top matches that of the conventional twin Higgs.

We estimate the tuning induced by this threshold correction as
\begin{equation}
\Delta\equiv \frac{|\delta m_h^2|}{m_h^2}
\end{equation}
as indicated by the dashed blue lines in Fig.~\ref{fig:mt1}. In the limit where $M_Q=M_U=0$, the tuning reduces to 
\begin{equation}
\Delta \approx \frac{f^2}{2 v^2}\approx 5
\end{equation}
as in the conventional twin Higgs. Here we have used that the fact that the SM quartic arises predominantly from the $\mathbb{Z}_2$-preserving, $SU(4)$-breaking radiative correction $\delta\sim \frac{3y_t^4}{16\pi^2}\log(y_t^2 f^2/\Lambda^2)$ \cite{Chacko:2005pe}. (See also section 3 of \cite{Craig:2015pha} for a detailed discussion.) We further observe that $\Delta$ is a rather mild function of $M_Q$ and $M_U$, and that even for $M_Q\sim M_U\sim$ 500 GeV, the tuning only increases by roughly a factor of two with respect to the conventional twin Higgs.

\subsection{Three-generation model\label{sec:generic}}
In the three-generation model, the twin sector has the same matter content as in the Standard Model, but with vector-like fermions. The Lagrangian is then  
\begin{equation}
\begin{split}
\mathcal{L}\supset &\; Y_U H' q' u' + Y_D H'^\dagger q' d' +Y_E H'^\dagger \ell' e' \\
&+ M_Q q'\bar q' + M_U u'\bar u' + M_D d'\bar d'+ M_L \ell'\bar \ell'+ M_E e'\bar e'\,,
\end{split}
\end{equation}
where all fermions carry the same quantum numbers as their Standard Model counterparts, but under the twin $SU(3)'\times SU(2)'$ rather than the SM gauge group. (With the exception that twin hypercharge is absent.) The relative magnitudes of all Yukawa's, except the top Yukawa, are in principle arbitrary, provided they are all much smaller than one. 
For simplicity, in this section, we will set all three twin Yukawa matrices equal to those in the Standard Model. 
As a final simplifying assumption, we also largely decouple the members of the $\mathbf{5}$-$\mathbf{\overline5}$ multiplets ($d', \ell'$) by setting $M_D\sim M_L\gg M_Q,M_U,f$. The twin leptons are therefore either decoupled or sterile and we do not further discuss them here. However as we will see, the $d'$ still have a role to play, as they induce flavor-changing higher dimensional operators. 

In the absence of the Yukawas and mass terms, the residual twin sector quarks then have a large flavor symmetry
\begin{equation}
U(3)_Q\times U(3)_U\times U(3)_D\times \overline{U(3)_Q}\times \overline{U(3)_U}\times \overline{U(3)_D}
\end{equation} 
which is maximally broken by the flavor spurions $Y_U$, $Y_D$, $M_Q$, $M_U$ and $M_D$. To preserve naturalness, we require $M_{Q,U}\lesssim 500$ GeV. 

As in the minimal vector-like model, the mass eigenstates are mixtures of the $SU(2)$ doublet and singlet quarks.  Consequently the $Z'$ generically has flavor off-diagonal couplings, which are large in the up sector. We will refer to this type of interaction as `twin flavor changing neutral currents' (twin FCNC's). Moreover it is generally also impossible to diagonalize the mass and Yukawa matrices simultaneously,  so we also expect large twin FCNC's in the Higgs sector.\footnote{Since the two sectors communicate exclusively through the Higgs portal, the presence of twin sector FCNC's does not imply a new sources of SM flavor violation.  SM flavor violation could in principle be induced by irrelevant operators, from integrating out the heavier states comprising the UV completion. (See \cite{Csaki:2015gfd} for a recent analysis in the context of the composite twin Higgs.) We will discuss this briefly when we turn to explicit constructions.}  
Even if we neglect the twin charm and up quark Yukawas, so that the eigenvalues of the up-type Yukawa matrix can be approximated by $\{y_t,0,0\}$, diagonalizing the $M_Q$ and $M_U$ matrices still leaves the up-type Yukawa matrix completely mixed. The presence of non-zero charm and up Yukawa couplings then has little additional effect. 
Therefore, each of the six mass eigenstates $u'_i$ contains a certain admixture of the top partner ({\it i.e.}, the one up-type state that couples strongly to the twin Higgs doublet).  
If we take $M_Q$ and $M_U$ to have eigenvalues of order $\overline{M} \ll y_t f$, as required for the vector-like twin Higgs mechanism to work, then there will be one heavy mass eigenstate $u'_6$ with mass   $\gtrsim y_tf/\sqrt{2}$, one light state $u'_1$ with a mass of order $\overline{M}^2/(y_t f)$, and four other states with mass of order $\overline{M}$.  
Specifically, if we take $\overline M$ in the 100--300 GeV range and $f\sim 3v$, we expect at least one state below 100 GeV and one around 750 GeV, similarly to the minimal vector-like model, plus four more scattered in between. 
In this scenario, typically only the heavy state $u'_6$ couples strongly to the Higgs sector.
The coupling of the lightest mass eigenstate to the Higgs is then slightly smaller than what it was in the minimal model, by up to a factor of $\sim 2$, because of the mixing with other light twin quarks.  

Since we took $M_D\gg M_Q$, the lowest mass eigenstates in the down sector $d'_1,d'_2,d'_3$ lie essentially at the same scale of the eigenvalues of $M_Q$, up to small corrections.  
These corrections, though small, induce $Z'$-mediated flavor changing interactions. 
Moreover, as for the up-sector, $Y_D$ generally has sizable off-diagonal entries in the mass eigenbasis, even if we only turned on its $y_b$ diagonal coupling. 
Explicitly integrating out the $d'$ results in the operator 
\begin{align}\label{eq:downFCNC}
 \frac{1}{2}v h\sum_{ij} c_{ij}\left[M_{Q,i} q'_j \bar q'_i + M_{Q,j} q'^\dagger_i \bar q'^\dagger_j \right]  \quad \mathrm{with} \quad c_{ij}\equiv \left(Y_D^\dagger \frac{1}{M_D^2} Y_D\right)_{ij}
\end{align}
and $M_{Q,i}$ the eigenvalues of $M_Q$. This induces a twin flavor changing interaction with the Standard Model Higgs, which can potentially be of phenomenological importance in some corners of the parameter space. (A similar higher dimensional operator may exist in the minimal vector-like model; however in that case it does not have any particular phenomenological significance.)

\section{Collider Phenomenology} \label{sec:pheno}

We now investigate the collider phenomenology of the two limits of the vector-like twin Higgs that we discussed in the previous section.  We will first discuss the hadrons of the twin sector, and then turn to how these hadrons may be produced through the Higgs portal, either by the decays of the 125 GeV Higgs $h$ or the radial mode (heavy Higgs) $\tilde h$.

\subsection{Twin Hadrons}
\label{subsec:twinhad}

We begin by reviewing the twin hadrons that arise in the fraternal twin Higgs of \cite{Craig:2015pha}, to which the reader is referred for further details.  In this model, there are two twin quarks,\footnote{In this paper, twin fields and parameters with a hat (e.g.~$\hat b,\, \hat m_b$) are those of the fraternal model discussed in \cite{Craig:2015pha}. Twin matter fields in the vector-like model, the main subject of the current paper, are denoted by primes (e.g.~$u', d'$). For the twin electroweak bosons $W', Z'$ and the confinement scale $\Lambda_c'$ there is no ambiguity, and they are denoted with a prime in both models.}  a heavy twin top partner $\hat t$ and a lighter twin bottom $\hat b$ with mass $\hat m_b = \hat y_b f /\sqrt{2}\ll f$.
There are also twin leptons $\hat \tau,\hat \nu$. The $\hat \tau$ must be light compared to $f$, and in the minimal version of the model, $\hat\nu$ is assumed to be very light.  There are three different regimes.
\begin{itemize}
\item
  If the twin confinement scale $ \Lambda'_{c} \ll \hat m_b$, the light hadrons of the theory are glueballs. The lightest glueball is a $0^{++}$ state $G_0$ of mass $m_0\sim 6.8 \Lambda'_{c}$.  $G_0$ can mix with $h$ and decay to a pair of SM particles.  Its lifetime, a strong function of $m_0$, can allow its decays to occur (on average) promptly, displaced, or outside the detector \cite{Juknevich:2009gg,Juknevich:2009ji}. (See \cite{Curtin:2015fna,Csaki:2015fba,Clarke:2015ala,Buckley:2014ika,Cui:2014twa} for detailed collider studies.) Most other glueballs are too long-lived to be observed, except for a second $0^{++}$ state, with mass $(1.8-1.9)m_0$, that can also potentially decay via the Higgs portal.   In addition there are twin quarkonium states made from a pair of twin $\hat b$ quarks. In this regime they always annihilate to glueballs.
\item
  Alternatively, if $m_0>4 \hat m_b$, then the glueballs all decay to quarkonium states.  Among these is a set of $0^{++}$ states $\hat \chi$.  (The lightest quarkonium states are $0^{-+}$ and $1^{--}$, so the $\hat \chi$ states are may not be produced very often.) The  $\hat \chi$ states can potentially decay via the Higgs portal and could decay promptly, displaced, or outside the detector.  However, twin weak decays to very light twin leptons, if present, can often short-circuit the Higgs portal decays, making the $\hat \chi$ states invisible.
\item
  In between, both $G_0$ and $\hat \chi$ can be stable against twin QCD decays, in which case they can mix.  The state with the longer lifetime in the absence of mixing tends, when mixing is present, to inherit the decay modes of (and a larger width from) the shorter-lived state.
\end{itemize}
Heavier states decay as follows: $W'\to \hat \tau\hat \nu$, $Z'\to\hat b\bar {\hat b},\hat\tau^+\hat\tau^-,\hat \nu\bar {\hat \nu}$, and $\hat t \to \hat bW'$.

\begin{figure}[h!]
  \includegraphics[width=0.48\textwidth]{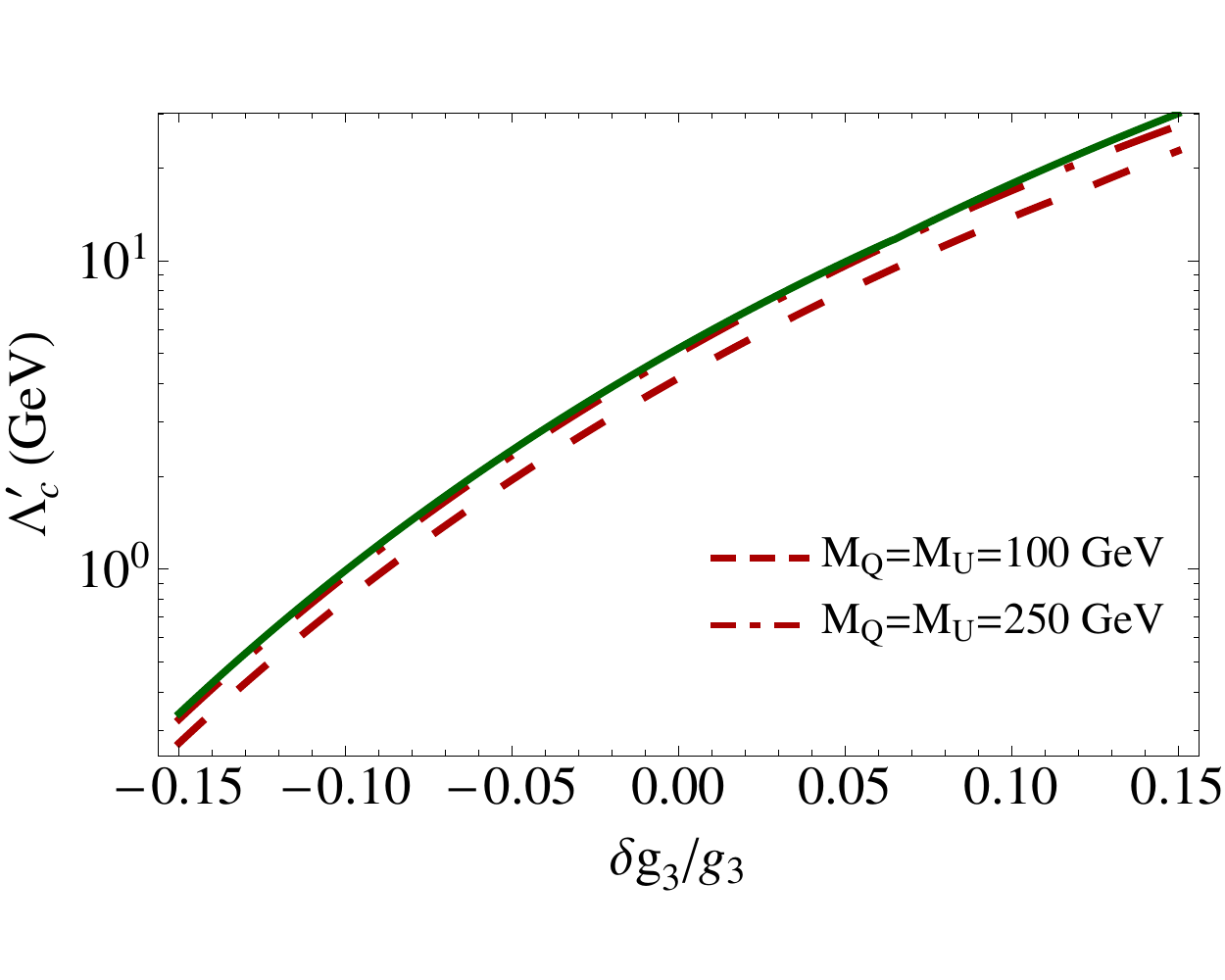}\hfill
  \includegraphics[width=0.48\textwidth]{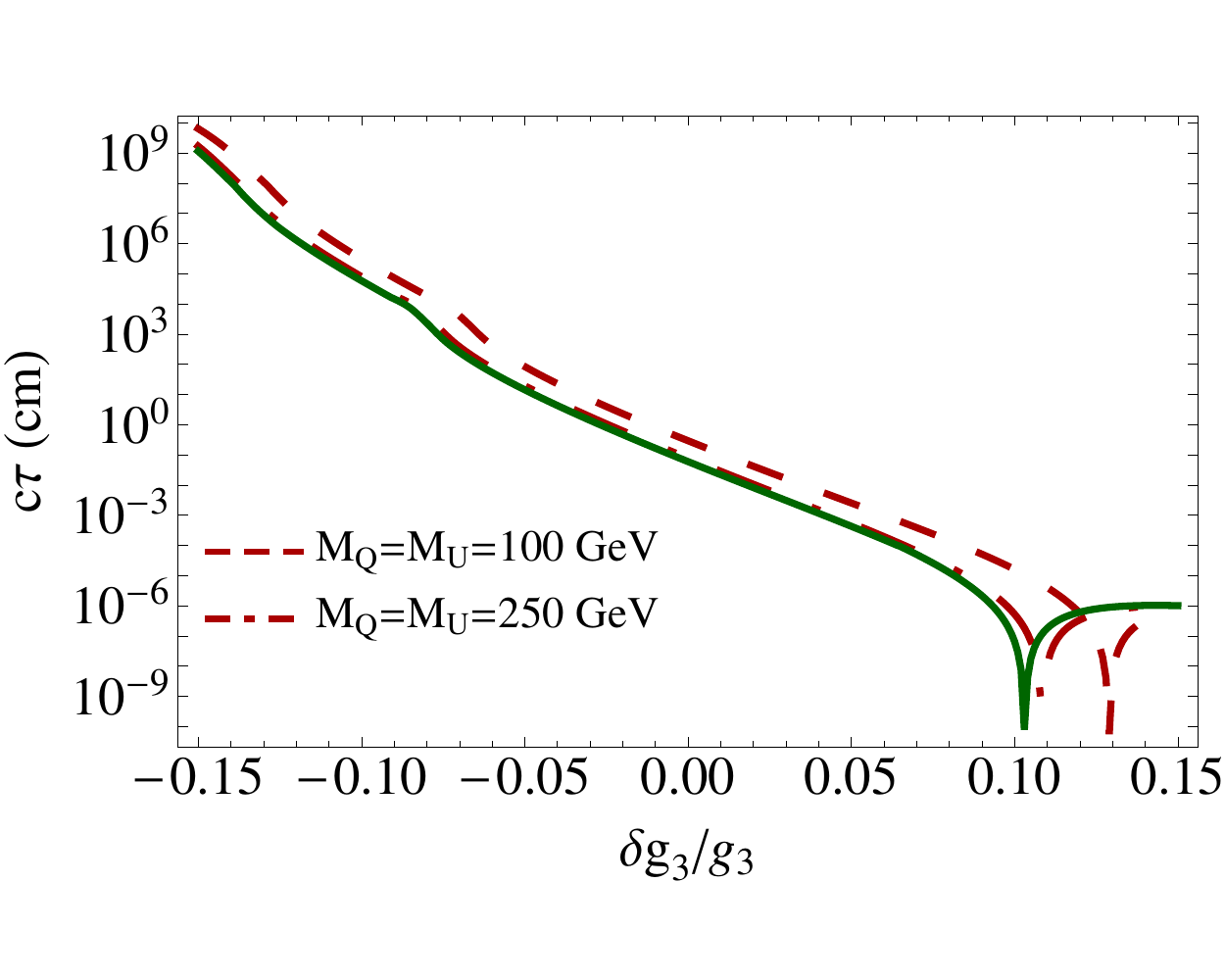}
  \caption{\label{fig:fratconf} Plots of the confinement scale $\Lambda'_{c}$ and $G_0$ glueball lifetime $c\tau$ as a function of the relative deviation $\delta g_3/g_3$ of the twin QCD coupling from the SM QCD coupling at the cut-off scale $\Lambda=5$ TeV. Shown are the fraternal case (solid green) and the minimal vector-like twin Higgs (dashed red). The RGE's were obtained with the SARAH package \cite{Staub:2013tta}. The confinement scale is defined as in \cite{Craig:2015pha}. The dip in $c\tau$ occurs when $m_0\sim m_h$.}
\end{figure}

The minimal model of the vector-like twin Higgs is remarkably similar to the fraternal twin Higgs, despite the fact that it has three twin quarks $t'_1,b'_1, t'_2$.  The surprise is that, as we saw in (\ref{eq:t1coupling}), the $t'_1$'s couplings to the Higgs are the same as for the twin $\hat b$ in the fraternal case, up to a minus sign and small corrections.  
The $b'_1$ itself plays a limited role for the light twin hadrons because its coupling to the Higgs is absent or at worst suppressed, as in (\ref{eq:downFCNC}).  Consequently the glueball phenomenology, and that of the $t_1'\bar t_1'$ quarkonium states, is very similar to that of the fraternal twin Higgs.  One minor effect (see figure \ref{fig:fratconf}), relevant only for low values of $M_Q$, is that the $b_1'$ makes the twin QCD coupling run slightly slower, so that
$ \Lambda'_{c}$ and $m_0$ are reduced by up to 20\%.  The relation between $m_0$ and the $G_0$ lifetime is the same as in the fraternal twin Higgs, so the lifetime correspondingly increases by up to an order of magnitude. This makes displaced glueball decays slightly more likely, as shown in the right-hand panel of figure \ref{fig:fratconf}. Here we took $|\delta g_3/g_3|<0.15$, which roughly corresponds to a fine tuning no worse than 30\%.

The significant new features in the minimal vector-like model  are consequences of 
the absence of light twin leptons, the role of $t_2'$-$t'_1$ mixing and the presence of the $b_1'$ in some decay chains.
\begin{itemize}
\item Without the twin leptons, $t_1' \bar t_1'$ quarkonium states cannot decay via twin weak interactions, so when the quarkonia are light compared to glueballs, the $\chi'$ states can only decay visibly, through the Higgs portal. (See Appendix A.2 of  \cite{Craig:2015pha}.)
\item Without light twin leptons, the $W'$ will be stable (and a possible dark matter candidate \cite{Garcia:2015loa}) if $W'\to \bar b_1' t_1'$ is closed.
\item Typically the $t_2'$ would decay to $b_1' W'$ and from there to $b_1'\bar b_1' t_1'$.  
However, this decay may be kinematically closed, and there is no twin semileptonic decay to take its place.  It therefore may decay instead via $t_2'\to t_1' Z' \to t_1' t_1' \bar t_1'$ or $t_1' h$, via equations (\ref{eq:yt1t2})-(\ref{eq:yt2t1}).

\item Because of twin hypercharge conservation, the $b_1'$ is stable if the decay $b_1'\to t_1' W'$ is kinematically closed, so there are also $b_1'\bar t_1'$ bound states.  Once produced, these ``flavor-off-diagonal quarkonia'' cannot annihilate and are stable. Flavor-diagonal bottomonium states annihilate to glueballs and/or, if kinematically allowed, toponium states.

\end{itemize}

Before moving on, let us make a few remarks about the behavior of quarkonium states, specifically in the limit where the glueballs are light.  When a twin quark-antiquark pair are produced, they are bound by a twin flux tube that cannot break (or, even when it can, is unlikely to do so), because there are no twin quarks with mass below the twin confinement scale.  The system then produces glueballs in three stages: (1) at production, as the quarkonium first forms; (2) as the quarkonium relaxes toward its ground state (it may stop at a mildly excited state); and (3) when and if the quarkonium annihilates to glueballs and/or lighter quarkonia.  During this process unstable twin quarks may decay via twin weak bosons, generating additional excited quarkonium states.  Obviously the details are very dependent on the mass spectrum and are not easy to estimate. The general point is that the creation of a twin quark-antiquark pair leads to the production of multiple glueballs, with potentially higher multiplicity if the quarkonium is flavor-diagonal and can annihilate.

\begin{figure}[t]\centering
  \includegraphics[width=0.48\textwidth]{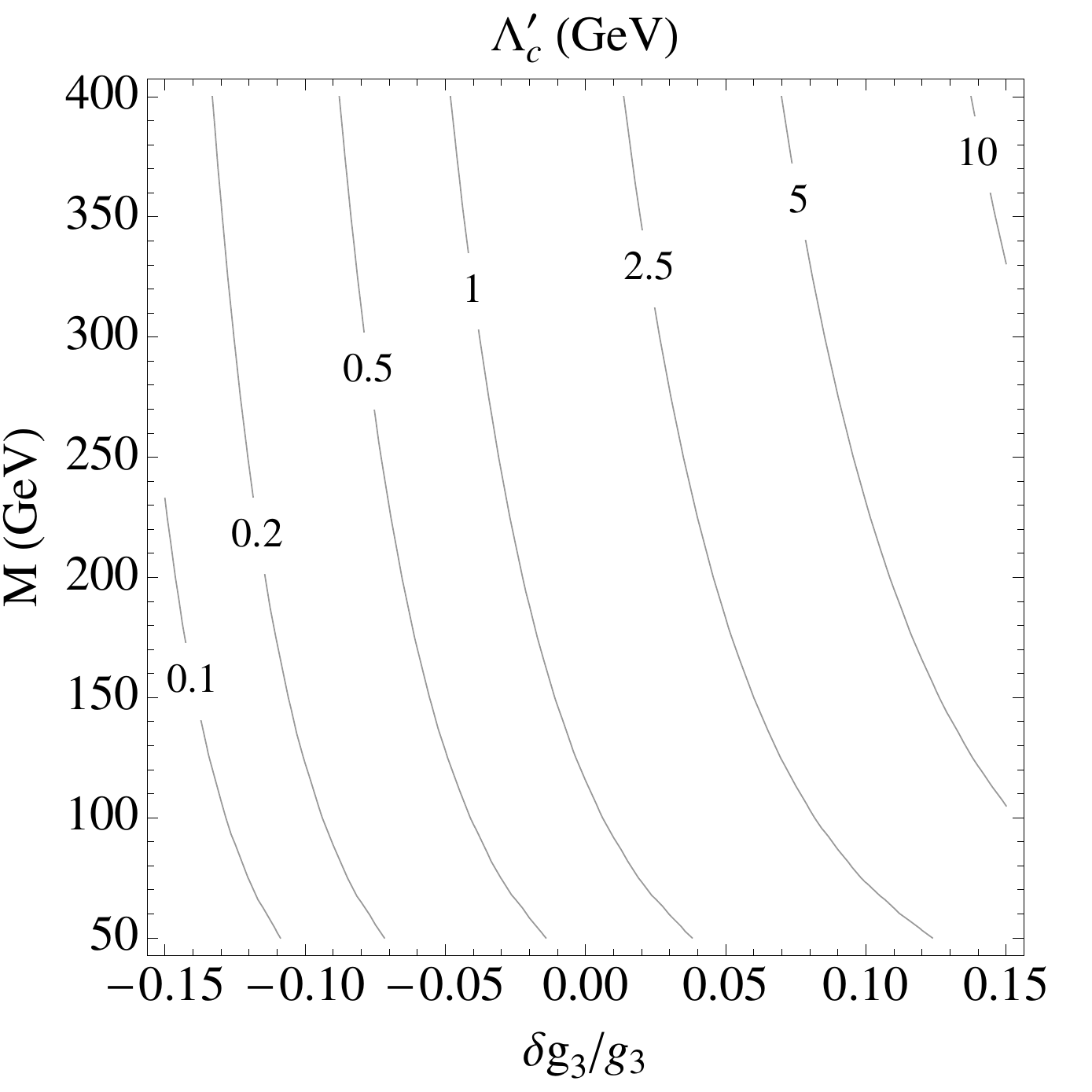}\hfill
  \includegraphics[width=0.48\textwidth]{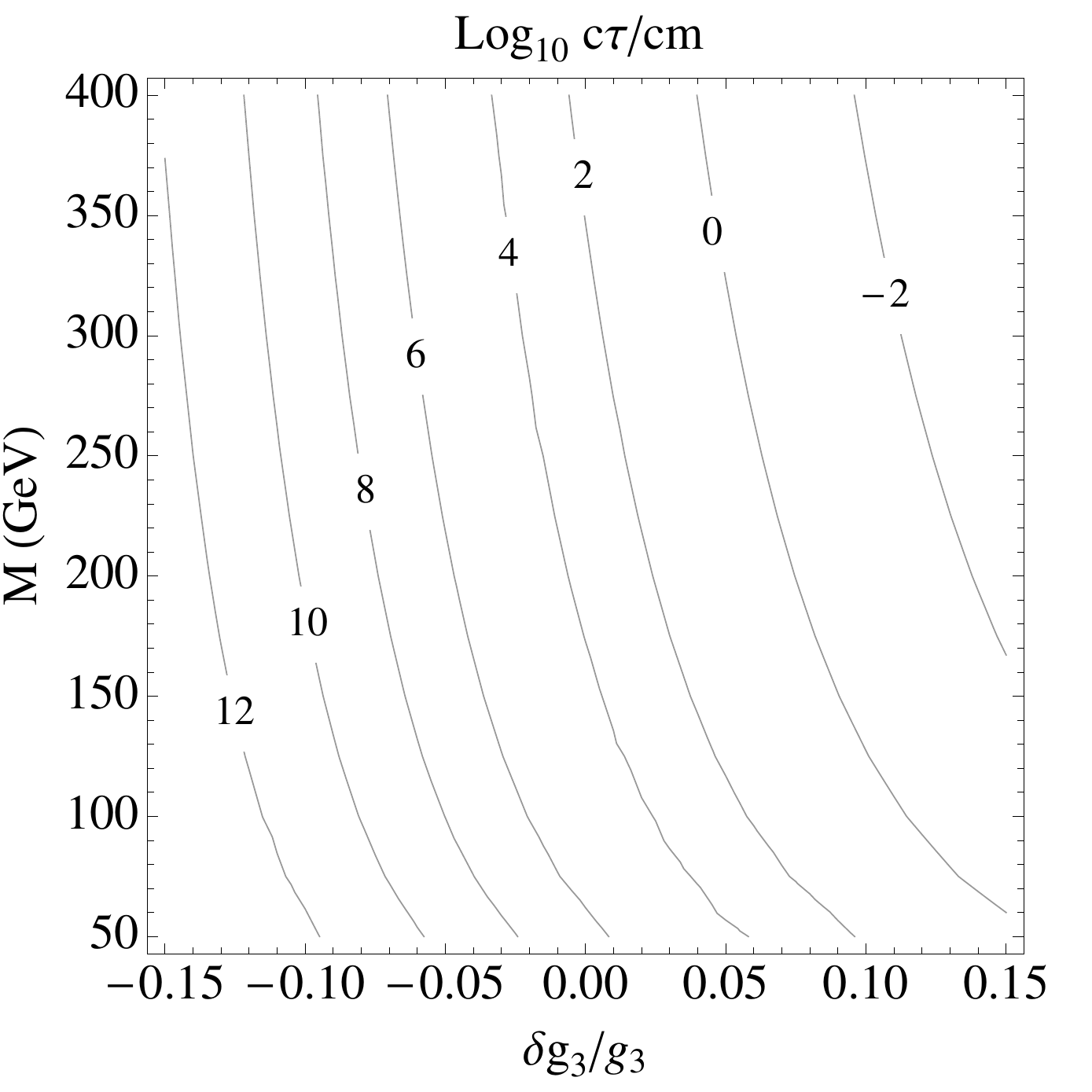}
  \caption{Twin confinement scale $\Lambda'_c$ and glueball lifetime $c\tau$ as a function of the vectorlike mass $M$ and a shift $\delta g_3/g_3$ in the twin QCD gauge coupling relative to the SM QCD coupling, at the cut-off $\Lambda=5$ TeV. Here we have taken $M_Q=M_U= M\times \One_3$. \label{confinement}}
\end{figure}

Let us turn now to the three-generation model, with its up-type quarks $u'_1,\dots, u'_6$ and down-type quarks $d'_1,\dots d'_3$ (plus three $SU(2)$ singlet down-type quarks with mass $\gg f$).
The most important difference from the fraternal twin Higgs is a twin QCD beta function that is less negative, which implies a lower confinement scale $\Lambda_c'$.  The twin glueball masses are therefore low and the lifetimes long, as shown in figure \ref{confinement}. For $\delta g_3<0$, the typical $G_0$ decays outside the detector.  Thus although the lower mass implies glueballs may be made in greater multiplicity, it may happen that few if any of the $G_0$ glueball decays are observable.  We also expect generally to be in the regime where the glueballs are the lightest states and flavor-diagonal quarkonia can annihilate into glueballs, so we expect no $\chi'$ 
decays to the SM.  As in the minimal vector-like model there are two stable twin quarks (here called $u'_1,d'_1$) and there can be flavor-off-diagonal $d'_1\bar u'_1$ quarkonia, which cannot annihilate.  However, heavier $d'_j$ quarks can in some cases be very long lived, with potentially interesting consequences.

\def\onoff{{(*)}}
Heavy twin $u_i$ quarks can decay via $W'^\onoff$, $Z'^\onoff$ or $h^\onoff$, and will cascade down to  $u'_1$ or $d'_1$. (The$\phantom{a}^{(*)}$ superscript indicates that the corresponding state may be on-shell or off-shell.)  Heavy $d_i$ quarks can decay via a $W'^\onoff$ if kinematic constraints permit. Heavy $d_i$ decays through $Z'^{(\ast)}$ or  $h^{(\ast)}$ are in principle possible as well, but are heavily suppressed. Since twin FCNCs are large, there can be  competition between the various channels, depending on the details of the spectrum.  Note that every $W'^\onoff$ or $Z'^\onoff$ in a cascade produces a new $q'\bar q'$, and thus increases the number of quarkonia by one.

\subsection{Production of twin hadrons via $h$ decays}

In the fraternal twin Higgs, as detailed in \cite{Craig:2015pha}, the rates of twin hadron production, and the decay patterns of the twin hadrons, depend on the confinement scale and the twin bottom mass.  Twin hadrons are produced in $h$ decays to twin gluons and/or twin $\hat b$ quarks.  The former is almost guaranteed but has a branching fraction of order $10^{-3}$. Of course the latter is forbidden if $\hat m_b > m_h/2$, but if allowed has a rate that grows with $\hat m_b\propto \hat y_b$ and easily dominates over decays to twin gluons.  In fact the rate is so large 
that corrections to $h$ decays exclude the model if $\hat m_b \gg 1.25 (f/v) m_b$  \cite{Craig:2015pha}. (See \cite{Craig:2015pha} for a discussion of important non-perturbative subtleties for $\hat m_b\sim m_h/2$.)

The minimal vector-like model is quite similar to the fraternal twin as far as $h$ decays. As in the fraternal model, there is a region excluded by an overabundance of $h \to t'_1 \bar t'_1$ decays, shown in the grey shaded region of figure 2, though this is a perturbative estimate with very large non-perturbative uncertainties at the upper edge.   The most important difference, as mentioned above, is that without light twin leptons, the $\chi'$ quarkonium states are more likely to decay visibly, making an experimentally accessible signal more likely.

In the three-generation model, the $u_1'$ coupling to the Higgs may vary by a factor of two or more compared to the minimal vector-like case, as a result of mixing with the other $u_i'$ states.  This changes Br($h\to u_1'\bar u_1'$) for a fixed $u_1'$ mass, and therefore also changes the range of $u_1'$ masses excluded by Higgs coupling measurements (the grey band of figure 2).

Since the less negative beta function of the three-generation model pushes down the glueball masses (see figure \ref{confinement}), in most of parameter space $u_1'\bar u_1'$ quarkonia will annihilate to glueballs. In some regimes 
  $G_0$ is  very light and long-lived; if $m_0<10$ GeV, $G_0$ decays are to $c\bar c, \tau^+\tau^-$ and the $G_0$ lifetime approaches the kilometer scale. All Higgs decays might thus be invisible.
  But more optimistically, small $m_0$ implies glueball multiplicity can be large.  With enough events and enough glueballs per event, 
  we may hope to observe Higgs decays to missing energy plus a single $G_0$ displaced decay, giving a low-mass vertex with a small number of tracks.  (Note that the vertices are distributed evenly in radius in this long-lifetime regime.)  This offers a challenging signal which pushes somewhat beyond what the LHC experiments have attempted up to now.

There is also a small possibility of observing off-shell Higgs bosons in $h$ decay. There is a region of parameter space where $h\to u_2'\bar u_1'$ is possible, followed by a prompt $u'_2\rightarrow u'_1 Z'^\ast\rightarrow u'_1\bar u'_1 u'_1$ or $u'_2\to h^\ast u_1'$ decay.
  If $m_{u_2} > 3 m_{u_1}$, the $Z'^\ast$ channel tends to dominate the decay; however if $m_{u_2} < 3 m_{u_1}$, then $u_2'\to h^\ast u_1'$ will proceed with 100\% branching fraction.

\subsection{Production of twin hadrons via the radial mode $\tilde h$}

The radial mode may be a relatively narrow resonance, if a linear sigma model describes the twin Higgs, or it may be wide and heavy if strong compositeness dynamics is involved.  If it is sufficiently light and/or wide,  $gg$ collisions at the LHC will be able to excite it.  For simplicity we will assume the mode is narrow and will refer to it as $\tilde h$, with mass $\tilde m$ that is not well-constrained but is likely in the 500-2000 GeV range.  The $\tilde h$ decays mainly to its Goldstone modes, namely the SM bosons $WW,ZZ,hh$ as well as the twin bosons $W'W',Z'Z'$, which may in turn decay to twin quarks.  Direct decays of $\tilde h$ to the twin quarks are possible though relatively suppressed, just as a heavy SM Higgs would decay rarely to fermions.

In the fraternal twin Higgs, $\tilde h$ decays to twin hadrons are most likely to occur through $\tilde h\to Z'Z'$, because the $Z'$ can decay to twin quarks with a branching fraction of order 60\%. The $W'$ decays only to $\hat\tau\hat \nu$ pairs.  Meanwhile $\tilde h$ decay to $\hat t$ pairs is highly suppressed by couplings and kinematics, but if it is present, the weak decay $\hat t\to \hat b\hat W$ leads to a single highly-excited twin bottomonium.  The bottomonium then deexcites as described in section \ref{subsec:twinhad}, typically producing multiple glueballs.

Without twin leptons and with both $t_1'$ and $b_1'$ quarks, the minimal vector-like twin Higgs differs from the fraternal twin in several ways.  Decays of $\tilde h$ to twin bosons may lead to many more twin hadron events, and higher multiplicity on average, because the $Z'$ always decays to $t_1'$ or $b_1'$ quark-antiquark pairs, and the $W'$ may be able to decay to $t_1'\bar b_1'$.  Each of these decays produces an excited flavor-diagonal or flavor-off-diagonal quarkonium.  Furthermore, the decay $\tilde h\to t_2'\bar t_1'$, though suppressed by a mixing angle, may be kinematically allowed even if $t_2'\bar t_2'$ is not.

Finally the $\tilde h$ decays in the three-generation model have the same rate as in the minimal model, but are potentially more diverse, possibly giving a new visible signature.  The more elaborate spectrum and large twin FCNCs allow $Z'\to u'_i\bar u'_j$ and $d'_k\bar d'_k$, and $W'\to u'_i\bar d'_k$ for $i,j =1,\dots,5$ and $k= 1,2,3$, depending on the spectrum of masses.  Also $\tilde h \to u'_6\bar u'_i$ may be possible though rare.  When $u'_i$ or $d'_k$ for $i,k>1$ is produced, a decay will ensue, possibly via a cascade, to $u'_1$ or $d'_1$.  These decays may produce an on- or off-shell $h$, 
as we now discuss.

Decays of the heavier $u'_i$ will most often go via $d'_k W'$ or $u'_j Z'$ if kinematically allowed, however decays to $h\, u_j'$ are also possible. This is especially so if the initial state is $u'_6$, which has sizable off-diagonal Yukawa couplings.
For lighter $u'_i$  the on-shell decays to $W'$ and $Z'$ are closed, so they are likely to decay via $u'_j h$ if kinematically allowed.  For $u'_i$ with mass less than $m_{u_1}+m_h$, the three off-shell decays via $W'^*,Z'^*, h^*$ all compete. If a decay mode to three twin quarks is open, decays through $W'^*$ and $Z'^*$ will typically dominate; otherwise the decay of the $u'_i$ must occur through an $h^\ast$.

Meanwhile, as discussed in section \ref{sec:generic}, see (\ref{eq:downFCNC}), the $d'_k$ have much smaller twin FCNCs.  The decay $d'_k\to W' u'_j$ always dominates if kinematically allowed.  Otherwise the decay $d'_k\to u'_1d'_1\bar u'_1$, via an off-shell $W'$, typically will dominate.  But for $d'_k$ too light even for this decay, only $d'_k\to h^\onoff d'_l$ may be available. The small FCNCs make this decay very slow, and in principle would even permit  observable displacement of the decay.  However, we must recall that each quark is bound to an antiquark and the quarkonium system relaxes to near its ground state. It seems likely, in this limit, that quarkonium relaxation and annihilation occurs before the individual quarks decay.

For flavor-diagonal $d'_k\bar d'_k$ quarkonia, $k>1$, annihilation occurs via twin QCD, and this is rapid. Flavor-off-diagonal quarkonia, including both $d'_k \bar d'_l$ and $d'_k \bar u'_1$, can only decay via twin electroweak processes, namely through flavor-changing exchange, in either the $s$- or $t$-channel, of a $W'$.    Still, this rate seems to exceed that of $d'_k$ decay.  With $m_q$ and $m_{\bar q}$ are the masses of the initial state quarks, an estimate of the annihilation width for a ground-state S-wave state to decay via a $W'$ is

\begin{equation}
  \Gamma\sim
  \alpha'^2_2\alpha'^3_3 \left(\frac{m_q+m_{\bar q}}{ M_{W'}}\right)^4  (m_q+m_{\bar q})
\end{equation}
times the squares of flavor mixing angles.   The rate is smaller for excited states, but the low glueball mass means that the quarkonium system is unlikely to get stuck in a highly excited state, so the suppression is not substantial.  Meanwhile this annihilation rate is to be compared with a decay such as $d'_k\to d'_l h$, which is two-body but suppressed by the coefficient $|c_{kl}|^2\sim y_b^4/M_D^2$ appearing in the operator (\ref{eq:downFCNC}), or a three-body decay via an off-shell $h$ which is suppressed by $y_b^6/M_D^2$.  The annihilation will have a much higher rate than the decay unless the relevant flavor mixing angles are anomalously small, the $d'_k$ and $d'_l$ are split by at least $m_h$, and $M_D\ll 5$ TeV, in which case the decay via an on-shell $h$ might be observable.   We conclude that for $d'_k$ that cannot decay via $W'^\onoff$, flavor-off-diagonal $d'_k\bar u'_1$ and $d'_k\bar d'_1$ quarkonia annihilate to lighter $d'_l \bar u'_1$, $u'_j \bar u'_1$ quarkonium states (plus at least one glueball). The $u_1\bar d_1$ quarkonium is stable. Again flavor-diagonal quarkonia annihilate to glueballs.

In sum,
the three-generation model offers cascade decays of heavier twin quarks which can generate additional
quarkonium states, along possibly with prompt on- or off-shell $h$ bosons from $u_i$ decay.
Consequently the final states from $\tilde h$ decay may have
\begin{itemize}
\item  twin hadrons (glueballs and flavor-off-diagonal quarkonia) that decay displaced or outside the detector;
\item prompt 
  on-shell $h$ decays;
\item prompt 
  decays of an off-shell $h$ to $b\bar b$, $\tau^+\tau^-$, or other jet pairs, similar to twin glueball final states but at a higher and variable mass.
\end{itemize}
Clearly, even with a very small rate for exciting the radial mode $\tilde h$, we should not overlook the possibility of a handful of striking events with substantial missing energy, at least one Higgs boson, and at least one displaced vertex with low mass.

\section{On the Origin of Symmetries} \label{sec:uv}
In the vector-like twin Higgs the $\mathbb{Z}_2$ symmetry is broken explicitly just by the presence of vector-like partners for the twin fermions. It is therefore essential to specify a UV completion from which the $\mathbb{Z}_2$ nevertheless emerges as an approximate symmetry in the IR. Such approximate IR symmetries often arise as a natural ingredient of orbifold constructions, making them ideal candidates for a UV completion of the vector-like twin Higgs. In the interest of clarity, we will first present a very simple and explicit 4D model based on the deconstruction of higher-dimensional theories \cite{ArkaniHamed:2001ca} with orbifold fixed points. 
These models possess the appropriate set of zero modes and the accidental $\mathbb{Z}_2$ symmetry. We will then discuss the relationship between these simple models and orbifold constructions.

\subsection{A simple UV completion} \label{sec:simpleuv}
\subsubsection{The model}
\begin{figure}[t]
\includegraphics[width=0.7\textwidth]{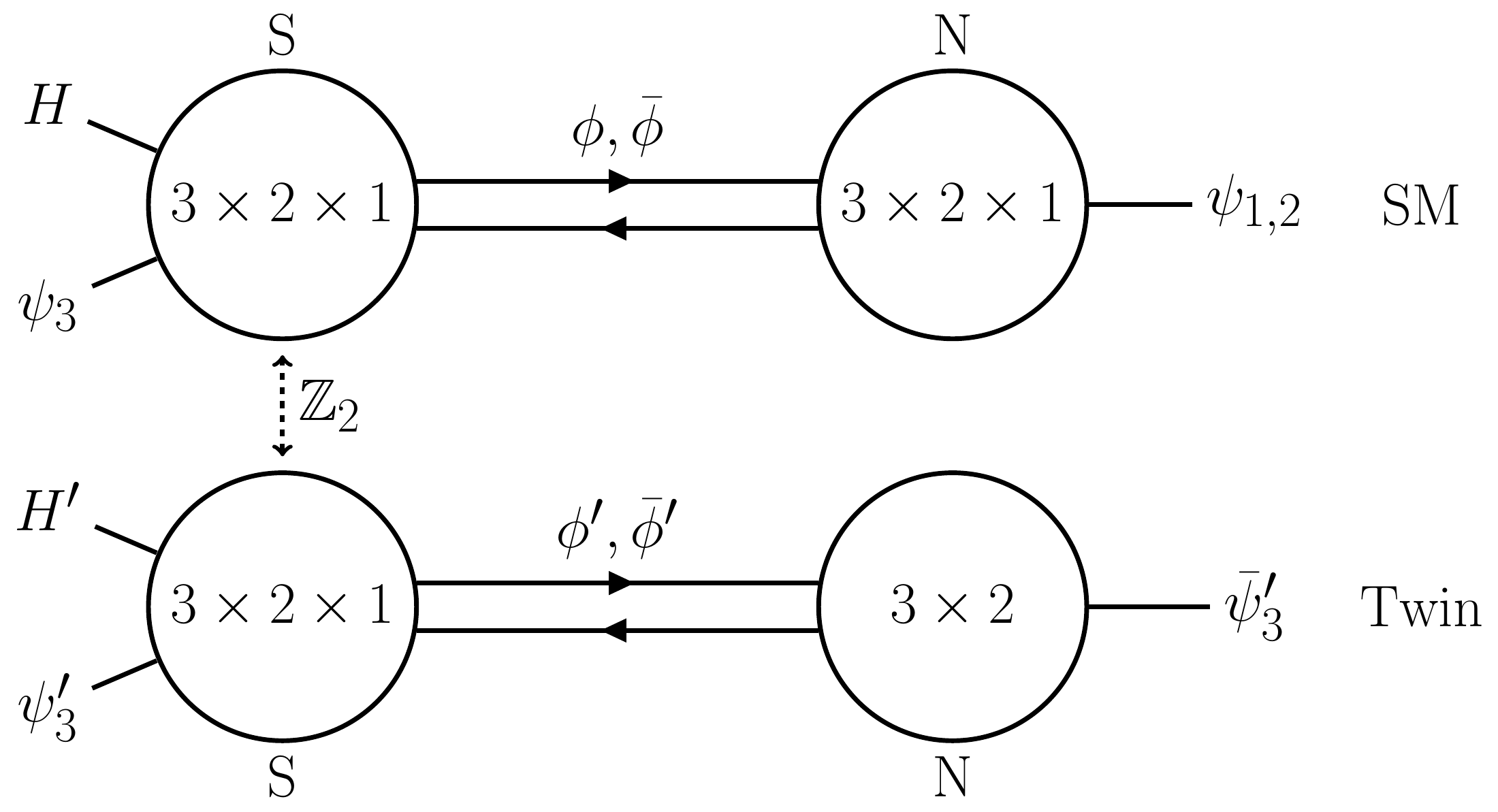}
\caption{\label{fig:mv} A schematic representation of the model. The $\psi_i$ ($\psi'_i$) each stand for a full generation of visible (twin) fermions, i.e. $\psi_i=(q_i, u_i, d_i, \ell_i, e_i)$ for $i=1,2,3$, and similarly for $\psi_3'$. The $\bar \psi'_3$ symbolizes a single anti-generation of twin fermions. There is an approximate permutation symmetry between the two $S$ nodes.}
\end{figure}

We begin with a simple UV completion for the vector-like twin Higgs that features the correct set of zero modes and an accidental $\mathbb{Z}_2$ symmetry. For concreteness, we focus on the minimal vector-like example, but the generalization to three generations in the twin sector is straightforward. Our example UV completion is heavily inspired by the dimensional deconstruction of an orbifold setup \cite{ArkaniHamed:2001ed,Csaki:2001em, Csaki:2001qm,Cheng:2001an, Kobayashi:2001fr, Falkowski:2002gx, Iqbal:2002ep, Cohen:2003xe, Dudas:2003iq, DiNapoli:2006kc} and shares many of its features. 
As indicated in Fig.~\ref{fig:mv}, the model can be divided into the SM and the twin sector, which each consist of a two-node quiver whose nodes are connected by a set of 
vector-like link fields, denoted $(\phi,\bar \phi)$ and $(\phi',\bar \phi')$ respectively. On the SM side each node contains a copy of the usual $SU(3)\times SU(2)\times U(1)$ gauge group, while on the twin side one node has the full $SU(3)\times SU(2)\times U(1)$ and the other only $SU(3)\times SU(2)$. On the latter node the $U(1)$ is present as a global symmetry, but it remains ungauged. The link fields organize themselves in complete ${\bf 5}$-${\bf \bar 5}$ multiplets of these gauge groups. We label the nodes in each sector by ``symmetric'' ($S$) and ``non-symmetric'' ($N$).
The $S$ node in on the SM side contains a SM Higgs field and a single, full generation of the SM fermions. Similarly, the $S$ node on the twin side contains a twin Higgs field and single generation of twin fermions. The $N$ node in the SM sector contains all the SM fermions from the first and second generations, while the $N$ node in the twin sector harbors a single twin anti-generation.
The SM and twin sectors only communicate with each other by means of the Higgs potential for $H,H'$ given in (\ref{daughertree}).

We further assume a $\mathbb{Z}_2$ permutation symmetry between the symmetric $S$ nodes of the two sectors, which ensures the presence of an approximate $SU(4)$ global symmetry in the Higgs potential.  The $\mathbb{Z}_2$ is only broken by the presence of the $N$ nodes on both sides. We assume all couplings of the link fields are moderate in size, such that their effects do not significantly violate the $\mathbb{Z}_2$ symmetry between the $S$ nodes.
In a more complete model, the $\mathbb{Z}_2$ symmetry of the $S$ nodes may arise from the unification of the SM and twin gauge groups into a single $SU(6)\times SU(4)$ node. While a detailed study is beyond the scope of the present work, as an intermediate step we provide a simple prescription for hypercharge in orbifold Higgs models in Appendix \ref{appHypercharge}. Constructions based on Pati-Salam unification or trinification are also possible \cite{Craig:2014aea}.

The SM Yukawa couplings to top, bottom, and tau, and the analogous couplings for their twin partners, are also present in the $S$ nodes, and the (approximate) $\mathbb{Z}_2$ symmetry assures they are (approximately) equal. The model is further equipped with the $SU(4)$-preserving and $SU(4)$-breaking quartics $\lambda$ and $\delta$, as in (\ref{twinlag}). The quartic $\lambda$ forms the only direct connection between the SM and twin sides of the quiver.

To address the ``big'' hierarchy problem (namely, the UV completion of the twin Higgs linear sigma model above the scale $\Lambda$), we take the theory to be supersymmetric down to a scale of order $\Lambda\sim 5-10$ TeV, much as in the supersymmetric twin Higgs \cite{Craig:2013fga,Chang:2006ra,Falkowski:2006qq}. As a consequence, it is natural to take the mass parameter $m^2$ in the Higgs potential to satisfy $m\sim \Lambda /4\pi$, such that the quartic $\lambda$ can be taken to be perturbative. The subtleties regarding the coset structure of strongly coupled models may therefore be bypassed \cite{Chacko:2005un,Barbieri:2015lqa}.  In addition we assume that the mechanism of supersymmetry breaking triggers vacuum expectation values for the link fields, such that both visible and twin sectors will see their $S$ and $N$ nodes Higgsed down to the diagonal $SU(3)\times SU(2)\times U(1)$ and  $SU(3)\times SU(2)$ respectively. (Twin hypercharge is fully broken.) The matter content in the visible sector is that of the Standard Model, while the twin sector contains a Higgs and a single vector-like generation. 

There are various options for generating a suitable link field potential that higgses each pair of $S$ and $N$ nodes down to the diagonal subgroup. The potential may be generated non-supersymmetrically, as in \cite{ArkaniHamed:2001vr}.  We here assume a set of soft-masses such that $\langle \phi\rangle\sim\langle \bar\phi\rangle\sim \Lambda \mathbb{1}$ and similarly for $\phi'$ and $\bar\phi'$. The $D$-term potentials for the link fields generate suitable quartics to stabilize the link fields at nonzero vev, provided that the soft masses satisfy some consistency conditions. (This is similar to what happens in the MSSM Higgs potential.)\footnote{The sole exception is a flat direction where the $SU(2)$ and $SU(3)$ components of the link fields acquire equal vevs, which may be stabilized by a $D$-term quartic for an additional gauged $U(1)$ under which $\phi, \bar \phi$ are vector-like; this gauge group may be broken at or above $\Lambda$. In this case some operators required for Yukawa couplings in the following subsection involve additional spurions for the $U(1)$ breaking.} Alternately, the link field potential may be generated supersymmetrically by including an additional singlet + adjoint chiral superfield on either the $S$ or $N$ nodes \cite{Cheng:2001an}.

The necessary Higgs potential is generated with a singlet coupling to the Higgses on each $S$ node as in \cite{Craig:2013fga}, and the potential (\ref{twinlag}) is reproduced in the decoupling limit where the additional states of the SUSY 2HDM are heavy. Note that SUSY provides a natural explanation for $\lambda >\! \!> \delta$, since $\lambda$ can be generated by a large $F$-term quartic while $\delta$ is generated by electroweak $D$-terms. For simplicity we will not commit to a specific model for supersymmetry breaking and mediation, save for enforcing the requirement that it respect the $\mathbb{Z}_2$ symmetry between the two $S$ nodes.

Finally, note that it is straightforward to modify this setup to accommodate a different set of zero modes. For example, we can obtain the three-generation model in section \ref{sec:generic} by simply putting three generations of matter fields on the $S$ nodes, as well as three anti-generations on the twin $N$ node. Another important example is that of the fraternal twin Higgs, which can be obtained by simply removing the $\bar\psi'_3$ from the quiver in figure \ref{fig:mv}.

\subsubsection{Mass scales\label{sec:massscales}}
The symmetry structure of the theory to some extent controls the form of Yukawa couplings. In particular, third-generation Yukawas are allowed at tree level since both the Higgses and third-generation fields are located on the symmetric node. However, the Yukawa couplings involving first two generations in the visible sector are forbidden by gauge invariance and instead must arise from irrelevant operators generated at a higher scale $\Lambda'$. In a supersymmetric theory these take the form
\begin{equation}
W \supset \frac{1}{\Lambda'}H_u {\phi_{D}} q_f  u_g + \frac{1}{\Lambda'^2}H_u \bar \phi_{T} \bar \phi_{D} q_f  u_3 \; +\text{ etc}
\end{equation}
with $f,g =1,2$. These operators may be induced by integrating out massive matter at the scale $\Lambda'$ as in \cite{Craig:2012hc}. The bi-fundamentals $\phi_D$  and $\phi_T$ are respectively the doublet and triplet components of the link field $\phi\equiv (\phi_T,\phi_D)$. When the link fields acquire vevs, this leads to Yukawa couplings with an intrinsic $\epsilon\equiv\langle \phi \rangle / \Lambda'\sim 0.1$ suppression. The resulting yukawa textures are
\begin{equation}
Y_U\sim \left(\!\begin{array}{ccc}\epsilon&\epsilon& \epsilon^2\\ \epsilon&\epsilon& \epsilon^2\\ \epsilon^2&\epsilon^2& 1\\\end{array}\!\right),\qquad Y_D\sim \left(\!\begin{array}{ccc}\epsilon&\epsilon& \epsilon^2\\ \epsilon&\epsilon& \epsilon^2\\ \epsilon&\epsilon& 1\\\end{array}\!\right),
\end{equation}
which can yield viable masses and mixings, though additional physics is required to explain the hierarchy between the first- and second-generation fermion masses.  Since these irrelevant operators are suppressed by the scale $\Lambda'$ and also may have small coefficients (indeed they cannot be too large or the $\mathbb{Z}_2$ will be badly broken), small Yukawa couplings for the first two generations result.  Flavor-changing effects that are not directly minimally-flavor-violating are present, since physics at the scale $\Lambda'$ generates flavor-violating four-fermion operators as well as effective Yukawa couplings. These flavor-violating operators are suppressed by both $\Lambda'$ and numerically small coefficients on the order of the CKM angles between the first two generations and the third generation, making it possible to accommodate flavor limits without further special alignment; see \cite{Craig:2012hc} for related discussion. Note that detailed flavor constraints may be relevant and perhaps even provide promising discovery channels; see \cite{Csaki:2015gfd} for a recent discussion of flavor signatures in UV complete twin Higgs models.

\begin{figure}[t]
\includegraphics[width=0.9\textwidth]{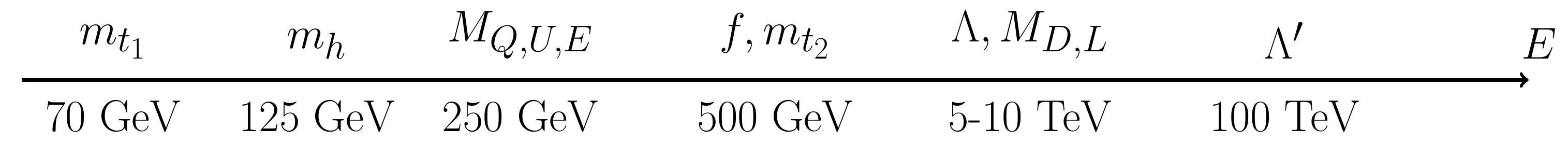}
\caption{\label{fig:mass} A schematic representation of the mass scales in the model. }
\end{figure}

Meanwhile, in the twin sector there are various possible marginal and irrelevant operators of interest, namely
\begin{equation}
  w_d\, d'{\bar\phi_{T}}'\bar d'+ w_\ell\, \ell'{\bar\phi'_{D}}\bar \ell'
  + \frac{w_q}{\Lambda'}  q'{\phi'_{T}\phi'_{D}}\bar q'
  +\frac{w_u}{\Lambda'}   u' \phi'_{ T} \phi'_{T} \bar u'
    + \frac{w_e}{\Lambda'}   e' {\phi'_{D} \phi'_{ D}}\bar e'
 \end{equation}
where $w_i$ are dimensionless coefficients. Once the link fields obtain $\mathcal{O}(\Lambda)$ vevs, the resulting mass spectrum has the following form:  
\begin{equation}
\begin{split}
&M_{D,L}\sim \Lambda\sim 5\; \mathrm{TeV}\\
&M_{Q, U, E} \sim \Lambda^2/\Lambda'\sim 250\; \mathrm{GeV}
\end{split}
\end{equation}
where for the latter estimate we take $\Lambda'\sim 100$ TeV. The twin neutrino, the left-handed twin tau and the right-handed twin bottom are therefore lifted, while the remaining states remain relatively light. The Yukawa-induced mixing between the left- and right-handed states is generally negligible for both for the bottom and the tau. Since the twin hypercharge is Higgsed at the scale $\Lambda$, the right-handed twin tau plays no role in the low-energy collider phenomenology of the twin sector.\footnote{The $e'$ could be a cosmogical issue; since its interactions with the rest of the twin sector are very weak, it could potentially overclose the universe. If this problem arises, it could be avoided if the reheating temperature is lower than $\Lambda$, or if the $e'$ can decay, either to $h \ell'$ if the spectrum permits, or through mixing with the SM neutrino sector, or through a dimension-six operator coupling $e'$ to twin quarks or SM fermions. }  Finally, note that $M_{q,u}\ll\Lambda$ automatically, as required by naturalness (see Section \ref{sec:vectortwin}). The twin tops are then heavily mixed, as discussed in Section \ref{sec:fraternal}. All mass scales are summarized in Fig.~\ref{fig:mass} for a benchmark point.

In order for the twin mechanism to be effective, the top Yukawa couplings of the twin and SM sectors should be equal to within about 1\%, while the twin and SM diagonal gauge couplings $g_{2,3}$ and $g'_{2,3}$ of the $SU(3)$ and $SU(2)$ groups should be equal to within about 10\% at the scale $\Lambda$.  Breaking of the $S$ and $N$ nodes to their diagonal subgroups will violate the latter condition unless the $N$ nodes of both the SM and twin sectors have couplings that either are nearly equal or are somewhat larger than the gauge couplings on the $S$ nodes. Expressed in terms of the coupling strengths $\alpha\equiv g^2/4\pi$, the $S$ nodes in each sector have a common $SU(3)$ coupling $\alpha_{3,S}$ while the $N$ nodes have relatively large (but generally unequal) $SU(3)$ couplings $\alpha_{3,N}$ and $\alpha'_{3,N}$. The couplings $\alpha_3, \alpha_3'$ of the unbroken $SU(3)$ gauge groups will then be equal up to corrections of order 
\begin{equation}\label{couplingtree}
\frac{\alpha_3-\alpha_3'}{\bar\alpha_3}=\frac{\bar\alpha_3}{\bar\alpha_{3,N}}\frac{(\alpha'_{3,N}-\alpha_{3,N})}{\bar \alpha_{3,N}}
\end{equation}
with $\bar \alpha_3=\sqrt{\alpha_3\alpha_3'}$, $\bar \alpha_{3,N}=\sqrt{\alpha_{3,N}\alpha'_{3,N}}$. {In addition there can be moderate one-loop threshold corrections proportional to $\log(\langle \phi \rangle /\langle \phi' \rangle)$.}
An analogous formula applies for  $SU(2)$. For instance, if $\alpha_{3,N}=2\alpha'_{3,N}$, the require accuracy can be achieved if $\alpha_{3,N}\gtrsim 0.38$ ($g_{3,N}\gtrsim 2.19$). With $\alpha_{2,N}=2\alpha'_{2,N}$, we need $\alpha_{2,N}\gtrsim 0.16$ ($g_{2,N}\gtrsim 1.4$). This implies that the $g_{2,N}$ coupling will reach a Landau pole before $10^6$ TeV, at which scale the model must be UV completed further.\footnote{Alternatively, we could have used $\mathbf{\bar3}$-{\bf3} $\oplus$ $\mathbf{\bar2}$-{\bf2} link fields, which removes the landau pole issue at the price of gauge coupling unification in the symmetric nodes.}  Thus we require the $N$ node gauge couplings be moderately large at the scale $\Lambda$.  We cannot allow them to approach $4\pi$, however, as would be the case at Seiberg fixed points; this would give $\phi,\bar\phi$ large anomalous dimensions, causing
unacceptable $\mathbb{Z}_2$-violating two-loop corrections to the couplings $\alpha_S$.

Having ensured an adequate degeneracy of the $SU(3)$ and $SU(2)$ couplings, we must also ensure that there are no additional large sources of radiative ${\mathbb{Z}_2}$-breaking which feed into the top yukawa. All third-generation yukawas are located on the $S$ nodes, and so do not pose a threat. The link fields cannot couple renormalizably to the top quarks because of their gauge charges.   The link fields may possess moderate ${\mathbb{Z}_2}$-breaking Yukawas to other fields, but these only feed into the running of the top yukawa at three loops, sub-dominant to the leading effect of the $SU(3)$ running.

\subsection{Connection with orbifolds}

Thus far we have presented a simple toy UV completion for the vector-like twin Higgs, but it is natural to wonder if a more general organizing principle might be at play. The key challenge in UV completing models like the fraternal or the vector-like twin Higgs is the fact that the twin sector looks radically different from the Standard Model sector, and the $\mathbb{Z}_2$ at best only persists as an approximate symmetry in a subsector of the theory. In previous work \cite{Craig:2014roa,Craig:2014aea}, we have shown that such approximate symmetries may be highly non-trivial and are a natural output of orbifold constructions. Concretely, one starts with a fully symmetric mother theory in the UV, which in our case would be a vector-like version of the Standard Model and a complete, vector-like twin copy. A suitable orbifold projection may then remove the unwanted degrees of freedom, while leaving behind a daughter theory with the desired accidental symmetry. Operationally the orbifold is carried out by identifying a suitable discrete symmetry of the theory and subsequently removing all degrees of freedom which are not invariant under the chosen discrete symmetry. In an actual model this projection can be implemented by selecting the zero modes of a higher dimensional theory, or by dimensional deconstruction. We first review the former, following \cite{Craig:2014roa},  and then provide a deeper motivation for the 4D model presented above.

\subsubsection{UV completion in 5D\label{sec:5duvcomple}}

We consider two copies of the MSSM gauge sector on $\IR^{4}\times S^{1}$, 
with a  global $\mathbb{Z}_2$ symmetry that sets the gauge couplings to be identical 
between the two. The theory further contains a whole vector-like third generation of MSSM matter multiplets. Owing to the fact that we start from a five-dimensional theory, the degrees of freedom within each multiplet resemble those of $4D$ $\CN=2$ theories from an effective four-dimensional viewpoint. Matter superfields in five dimensions descend to hypermultiplets in four; the latter can be conveniently thought of as a pair of chiral and anti-chiral $\CN=1$ superfields in the $4D$ effective theory. The matter fields are thus organized in terms of the hypermultiplets $\Psi_3=(\psi_3,\psi_3^c)$ and $\bar\Psi_3=(\bar\psi_3,\bar\psi_3^c)$, where the $\psi_3$ and $\bar\psi_3$ were defined in the caption of figure \ref{fig:mv}. The  $\psi_3^c$ and $\bar\psi_3^c$ are an additional set of fermion representations conjugate to $\psi_3$ and $\bar\psi_3$.
The matter content of the twin sector is identical, as required by the  $\mathbb{Z}_2$ symmetry. We denote it by the pair of hypermultiplets $\Psi_3'$ and  $\bar \Psi_3'$.

We take the $S^1/(\IZ_{2}\times \tilde {\mathbb{Z}}_2)$ orbifold of this mother theory: denoting spacetime coordinates $(\vec x,y)$, the action of the orbifold group on 
spacetime is the familiar (see for example \cite{Hebecker:2001wq})
\be \label{eq:spacetimeaction}
	P: y\to -y\qquad \tilde P: y\to \pi R - y\,.
\ee
The fundamental domain is thus $(0,\pi R/2)$, with $y=0$ being a $P$ fixed-point and $y=\pi R/2$ a $ \tilde P$ fixed point. We refer to these fixed points as the `symmetric' and `non-symmetric' brane respectively, for reasons that will become clear momentarily. 

$P$ and $ \tilde P$ also act on fields, in fact those fields which transform non-trivially under $P$ and/or $\tilde P$ must vanish at the corresponding orbifold fixed point(s), and their zero modes will be absent from the effective 4D theory.
The spacetime actions of both $P,\tilde P$ on superfields are identical: on the vector-multiplets they act by $(V,\Sigma)\to(V,-\Sigma)$, where $V$ and $\Sigma$ are the $\CN=1$ vector and chiral multiplets respectively. On matter hypermultiplets, the space-time action of $P,\tilde P$ takes  e.g.~$(\psi_3,\psi_3^{c})\to(\psi_3,-\psi_3^{c})$.
In addition to this, the $\IZ_{2}\times\tilde\IZ_{2}$ acts on the space of fields, with the following assignments:
We take $P$ to act trivially on the target space, while $\tilde P$ takes $\phi\to \tilde\eta_{\phi}\phi$ with $\eta_{\phi}=\pm1$. 
The combined action on the vector multiplets and the matter multiplets is given in the following table
\be
\begin{array}{c|c|c}
	& \text{Vector multiplet} & \text{Hypermultiplet} \\
	\hline
	P & (V,\Sigma)\rightarrow(V,-\Sigma) & (\phi,\phi^{c})\rightarrow(\phi,-\phi^{c}) \\
	\tilde P & (V,\Sigma)\rightarrow(\tilde \eta V,-\tilde \eta\Sigma) & (\phi,\phi^{c})\rightarrow(\tilde \eta\phi,-\tilde \eta\phi^{c}) 
\end{array}
\ee
where $\tilde\eta=\pm 1$ can be chosen for each individual field. The hypermultiplet ($\phi,\phi^c)$ can represent any of the matter hypermultiplets we introduced before.
In the language of the $4D$ $\CN=1$ superfields, only those which transform with a $(+,+)$ sign under $(P,\tilde P)$ can contribute a zero-mode to the effective 4D theory, since a negative sign under either operator requires the field to vanish at the corresponding brane.
In fact, the $P$ action manifestly breaks $\CN=2$ supersymmetry down to $\CN=1$: it requires both the $\Sigma$-component of all $5D$ vector multiplets the $\phi^{c}$ component of all $5D$ matter multiplets to vanish on the symmetric brane, thus killing the corresponding zero modes. 

On top of the supersymmetry breaking, $\tilde P$ further acts in the way specified by the following table\footnote{Strictly speaking, the $-$ condition for $A'^{(1)}$ does not correspond to an orbifold projection, however it is nevertheless self-consistent to impose a Dirichlet boundary condition on this field.}
\begin{equation}
\begin{array}{c|ccc|cc||ccc|cc}
&A^{(3)}&A^{(2)}&A^{(1)}  %
&\psi_{3}&\bar\psi_{3}  %
&{A'}^{(3)}&{A'}^{(2)}&{A'}^{(1)}   %
&\psi'_{3}&\bar\psi'_{3}\\
\hline
\tilde\eta%
&+&+&+   %
&+&-   %
&+&+&-   %
&+&+ 
\end{array}
\end{equation}
This implies a vanishing (Dirichlet) condition on the non-symmetric brane for certain $\CN=1$ components. In gauge fields the boundary condition applies to the $\Sigma$-component if $\tilde\eta=1$, or to the $V$-component if $\tilde\eta=-1$.
Overall, all $5D$ vector multiplets with $\tilde\eta=+1$ will descend to $4D$ $\CN=1$ vector multiplets, while ${A'}^{(1)}$ is entirely removed from the spectrum.
By analogous reasoning, all the $5D$ matter fields with $\tilde\eta=+1$ descend to $4D$ $\CN=1$ chiral multiplets, while $\bar\psi_{3}$ does not contribute zero modes to the $4D$ effective theory, since its components must vanish on either brane.

Finally, in each sector we introduce a pair of $4D$ $\CN=1$ Higgs multiplets $(H_u, H_d)$ and $(H'_u, H'_d)$, localized on the symmetric brane, along with a singlet chiral multiplet $S$. A $\IZ_2$-symmetric superpotential $W = \sqrt{\lambda} S (H_u H_d + H'_u H'_d)$ on the symmetric brane gives rise to the $SU(4)$-symmetric quartic $\lambda$, while $\IZ_2$-symmetric yukawa couplings connect these Higgses to the bulk fields.

The resulting $4D$ zero-mode spectrum includes a chiral copy of the MSSM and a vector-like copy of the twin sector, realizing a $5D$ supersymmetric UV completion of the vector-like twin Higgs. Our choice of boundary conditions leaves a zero-mode spectrum with unbroken $\CN = 1$ supersymmetry (in contrast with, e.g. folded SUSY \cite{Burdman:2006tz} where the boundary conditions break all supersymmetries). Further soft supersymmetry breaking may be introduced through local operators on the symmetric $y = 0$ brane, so that soft masses remain $\IZ_{2}$-symmetric.

It should be noted that bulk mass terms of the form $M \left(\psi_3 \bar \psi_3+\psi'_3 \bar \psi'_3\right)$  softly break the $\tilde {\mathbb{Z}}_2$ which we used for the orbifold. On the level of the zero modes, this is precisely the origin of the soft $\mathbb{Z}_2$ breaking by the vector-like mass terms, as discussed in section \ref{sec:vectortwin}. This procedure is easily generalized to a three-generation Standard Model, with all fermions in the bulk. Alternatively one may localize only a copy of the lowest Standard Model generations on the $ \tilde P$ brane.

While this model exemplifies the key features of a $5D$ realization of the vector-like twin Higgs, we note that it suffers a modest shortcoming related to the choice of a flat 5th dimension. In general, large brane-localized kinetic terms on the non-symmetric brane at $y = \pi R / 2$ will shift the effective $4D$ couplings of zero-mode states. The effect on SM and twin gauge couplings is  benign, but the shift in the SM and twin top yukawa couplings is typically larger than the percent-level splitting allowed by the twin mechanism. Such non-symmetric brane-localized terms can be rendered safe in a flat fifth dimension using bulk masses for third-generation fields of order $M \sim 1/R$ (thereby sharply peaking the corresponding zero mode profiles away from the non-symmetric brane), but at the cost of unreasonably large vector-like masses for the twin sector zero modes. Alternately, the theory may be embedded in a warped extra dimension where the bulk warp factor strongly suppresses the impact of non-symmetric brane-localized kinetic terms. The general features discussed in this section carry over directly to the warped case, although detailed model-building in a warped background is beyond the scope of the present work.\footnote{In contrast to holographic twin Higgs models \cite{Geller:2014kta, Barbieri:2015lqa, Low:2015nqa}, in this case the scale of the IR brane can be somewhat above the scale $f$, with supersymmetry protecting the linear sigma model. Thus it is sufficient for the accidental symmetry of the Higgs sector to be $SU(4)$ rather than $O(8)$, since higher-dimensional operators are parametrically suppressed \cite{Chacko:2005un}. }

\subsubsection{UV completion in 4D}
Finally, we come full-circle by presenting a 4D theory which yields the same spectrum as the 5D setup in the previous section, and illustrate the relation to our initial 4D model. The basic template for such a setup is a chain of `nodes' with the gauge group in the bulk of the 5D theory, connected by bi-fundamental link fields. To automatically cancel any gauge anomalies at the boundaries, we take the link fields to be vector-like.\footnote{Note that a literal deconstruction of the 5D theory would entail oriented, rather than vector-like, link fields with additional matter on the end nodes to cancel anomalies.  } The last node on one end of the chain contains the reduced gauge group of the daughter theory, which in our case is the same as the full bulk gauge theory, minus twin hypercharge. We call this node the `non-symmetric node', in analogy with the `non-symmetric brane' in the previous section. The node on the opposing end of the quiver has the full gauge symmetry plus the global $\mathbb{Z}_2$, and we will refer to it as the `symmetric node', again in analogy with the terminology in the previous section. When the link fields are Higgsed, this construction yields a spectrum identical to the KK-modes of the 5D gauge theory. 

\begin{figure}[t]
\includegraphics[width=\textwidth]{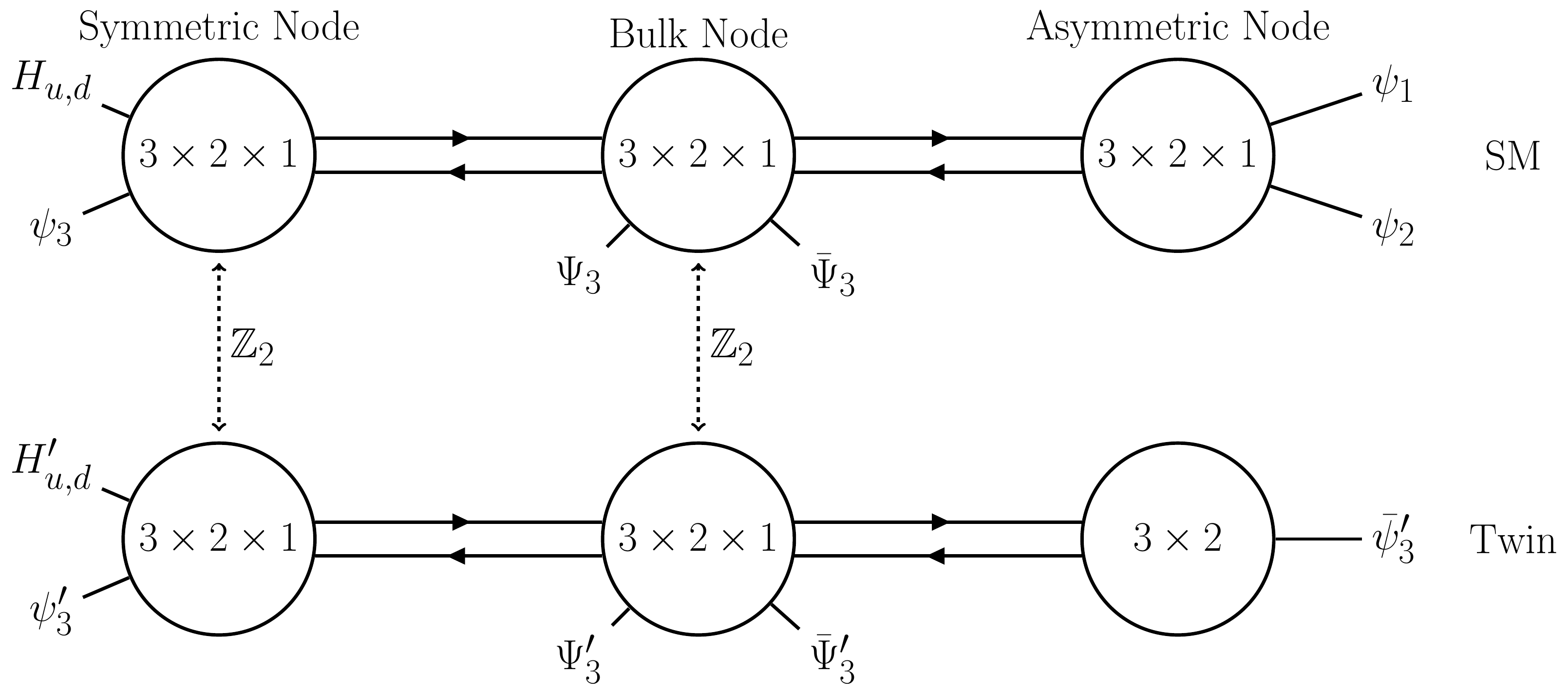}
\caption{\label{fig:deconstr} A schematic representation of the deconstruction of the orbifold model. For simplicity, only one bulk node is shown. The notation is as in Section \ref{sec:5duvcomple}. }
\end{figure}

The remaining matter content is specified according to the following rules:
\begin{itemize}
\item All fields which propagate in the 5D bulk appear on the bulk nodes. These correspond the matter hypermultiplets, introduced in the previous section.
\item Fields which have a zero mode in the 5D theory appear on one of the boundary nodes. Which boundary node they are attached to is a priori arbitrary, and all  multiplets on the boundary nodes are $\CN=1$ and chiral. Fields which do not have zero modes appear on neither boundary node. 
\end{itemize}
In our example, we choose to attach $\psi_{3}$ and $\psi_{3}'$ to the symmetric boundary node, and to move $\bar\psi_{3}'$ to the non-symmetric node. This has the advantage that the $\mathbb{Z}_2$ symmetry of the symmetric node is manifestly preserved. In analogy with the previous section, we also add the $H_{u,d}$ and $H'_{u,d}$ multiplets on the symmetric boundary node.
Neither $\bar\psi_{3}$ nor any of the anti-chiral components of the bulk hypermultiplets have a zero mode, and they therefore do not appear on the boundaries. This construction is shown schematically in figure \ref{fig:deconstr}.
 
The resulting quiver has a strong resemblance to the model of section \ref{sec:simpleuv}. In particular, we can obtain the quiver in figure \ref{fig:mv} by simply dropping all bulk nodes from the model. This removes all KK-modes from the model, and strictly speaking its interpretation in terms of the deconstruction of an extra dimension is lost. However since the KK-modes are likely to be out of reach at the LHC, the two options are likely indistinguishable in the near future. 
 
\section{Conclusions\label{sec:conc}} 

The tension between LHC null results and anticipated signals of conventional top partners motivates alternative theories of the weak scale with novel signatures. Many such alternative theories, including the twin Higgs and folded supersymmetry, exhibit hidden valley-type phenomenology intimately connected to the stabilization of the weak scale. In their simplest incarnations, these theories and their signatures are made rigid by the requirement of exact discrete symmetries. Far greater freedom is possible for both models and their signatures if the discrete symmetries are approximate, rather than exact. The precise signatures of these models depend, however, on both the detailed physics of the dark sector and the UV completion, which is required to justify the presence of approximate stabilization symmetries. 

In this paper we present an intriguing deformation of the twin Higgs model in which the twin sector may be vector-like without spoiling naturalness. From a bottom-up point of view, this deformation is innocuous in that the presence of these extra mass terms is merely a soft breaking of the twin $\mathbb{Z}_2$ and should therefore not reintroduce the quadratic sensitivity to the cut-off of the theory. However, while the vector-like mass terms represent a soft $\mathbb{Z}_2$ breaking, the presence of vector-like states constitutes a hard breaking (through, e.g., their impact on the running of couplings in the twin sector) that requires a UV completion. We show that this setup can be UV completed in the context of the orbifold Higgs and we provide an explicit model based on dimensional deconstruction. (A similar mechanism is at work in the Holographic Twin Higgs \cite{Geller:2014kta} where spontaneous breaking of a bulk symmetry leads to modest bulk masses for twin sector fermions.) The same mechanism can moreover be used as a UV completion of the fraternal twin Higgs.

The phenomenology of the vector-like twin Higgs is very rich, and depends strongly on the number of twin generations, the flavor texture of the vector-like mass terms and their overall size. In this paper we have analysed two example models where the twin quarks are all relatively heavy compared to the twin confinement scale. In this case, the collider phenomenology is similar to that of the fraternal twin Higgs, but with a few important differences. Due to the extra matter charged under twin QCD, the twin confimenent scale tends to be somewhat lower, which increases the likelihood for glueballs to decay displaced. Due to absence of light leptons, either the lightest state in the down-sector or the $W'$ is stable. However perhaps the most striking feature is the presence of order-one flavor changing neutral currents in the twin sector. As a result, cascade decays of heavier twin fermions may produce spectacular events with glueball decays in association with one or more on or off-shell Higgses.

There are a number of interesting future directions worth pursuing: 
\begin{itemize}
\item In this paper we have assumed a $[SU(3)\times SU(2)]^2$ gauge group and imposed the $\mathbb{Z}_2$ symmetry by hand on the symmetric nodes in figures \ref{fig:mv} and \ref{fig:deconstr}. In \cite{Craig:2014roa,Craig:2014aea} we showed how the $\mathbb{Z}_2$ symmetry can be an automatic ingredient if the Standard Model and twin gauge interactions are unified at some scale near 10 TeV. It may be worthwhile to investigate under what conditions it is possible to generalize this idea to the vector-like twin Higgs, and in particular to construct four-dimensional UV completions.
\item We also restricted ourselves to a broad-brush, qualitative description of the collider phenomenology. It would be interesting to study some well motivated benchmark scenarios in enough detail to get a quantitative idea about the reach of the LHC for these models. Of particular interest here would be the signatures resulting from the production of the radial mode or the lowest KK-states (if they are present), along the lines of \cite{Cheng:2015buv}.
\item A final direction for further progress is related to cosmology. While the traditional mirror twin Higgs requires a very non-standard cosmology to avoid CMB constraints on a relativistic twin photon and twin neutrino's, this tension can be relaxed significantly in the fraternal twin Higgs \cite{Craig:2015xla,Garcia:2015loa}. In the vector-like twin Higgs, this tension is removed entirely since the neutrinos are vector-like and can therefore be heavy. The lightest twin lepton may still be a twin WIMP dark matter candidate and its annihilation cross section and relic density now depends on the spectrum of the twin quarks. Alternatively, the $W'$ may be stable and could make up (part of) the dark matter \cite{Garcia:2015loa}. Another intriging possibility opens up when the twin quarks are light, as now the twin pions could be the dark matter and freeze out from the twin strong interactions through the SIMP mechanism \cite{Hochberg:2014dra,Hochberg:2014kqa}. Even if the CMB constraints can be avoided, this idea is still difficult to realize in the traditional mirror twin Higgs due to the number of light flavors required for the SIMP mechanism to operate. Both this issue and the CMB constraints can be naturally addressed in the vector-like twin if the vector-like masses are below the confinement scale.

\end{itemize}

\section*{Acknowledgments}
We thank Hsin-Chia Cheng, Tim Cohen, Csaba Csaki,  Michael Geller, Adam Falkowski, Roni Harnik, Yonit Hochberg, Eric Kuflik, Tim Lou, John March-Russell, Michele Papucci, Dean Robinson and Yuhsin Tsai for useful conversations.  
NC is supported by the Department of Energy under the grant DE-SC0014129. The work of SK was supported by the LDRD Program of LBNL under U.S. Department of Energy Contract No. DE-AC02-05CH11231. The work of PL is supported by the Carl Tryggers Stiftelsen. SK and MJS thank the Gallileo Galilei Institute for Theoretical Physics where part of this work was completed.

\appendix

\section{Hypercharge in Orbifold Higgs Models \label{appHypercharge}}
In \cite{Craig:2014roa} we presented a class of models where the twin Higgs or a generalization arises from an orbifold of theory where the SM and twin gauge groups are unified. The explicit unification of the gauge groups of both sectors then provides a natural explanation for the presence of the (approximate) $\mathbb{Z}_2$. However, in order to ensure that the twin sector is dark under SM hypercharge, these models tend to require (partial) low scale gauge coupling unification  of the SM gauge groups. This can be accomplished, for example, with an enlarged version of Pati-Salam unification or trinification.

Here we provide an alternative setup with a $\mathbb{Z}_2\times \mathbb{\tilde Z}_2$ orbifold where such low scale unification is not required. To illustrate the principle, we present a simple toy model which only includes the top and Higgs sectors. The generalization to a full model is straightforward. We consider an $SU(6)\times SU(4)\times U(1)_A\times U(1)_B$ gauge group and two sets of fields ($H_A, Q_A, U_A$) and ($H_B, Q_B, U_B$) with representations as in table \ref{tab:appendix}. We can identify $U(1)_A$ and $U(1)_B$ with SM and twin hypercharge respectively. The action is
\begin{equation}\label{apphyperlag}
-\mathcal{L}\supset y_t H_A Q_A U_A + y_t H_B Q_B U_B
\end{equation}
were we assume a $\mathbb{Z}_2$ symmetry which exchanges the $A\leftrightarrow B$. 

\begin{table}[h]
$$
\begin{array}{c|cc|cc}
& SU(6)&SU(4)&U(1)_A&U(1)_B\\\hline
H_{A}& 1 &\square&1/2&0\\
Q_{A}& \square &\overline\square&1/6&0\\
U_{A}& \overline\square &1&-2/3&0\\\hline
H_{B}& 1 &\square&0&1/2\\
Q_{B}& \square &\overline\square&0&1/6\\
U_{B}& \overline\square &1&0&-2/3\\
\end{array}
$$
\caption{Matter content of the mother theory. $A$ fields carry SM hypercharge, $B$ fields do not. \label{tab:appendix}}
\end{table}

\begin{table}[h]
$$
\begin{array}{c|cc|cc}
& SU(3)&SU(2)& SU(3)'&SU(2)'\\\hline
h_A& 1 &\square&1&1\\
q_A&\square&\overline \square&1&1\\
u_A&\overline\square&1&1&1\\\hline
h'_B&1&1& 1 &\square\\
q'_B&1&1&\square&\overline \square\\
u'_B&1&1&\overline\square&1\\\hline\hline
q_B&\square&\overline \square&1&1\\
q'_A&1&1&\square&\overline \square\\
\end{array}
$$
\caption{All fields surviving the $\mathbb{Z}_2$ projection. The fields below the double line are removed by the $\mathbb{\tilde Z}_2$ projection. The fields labeled with the $A$ subscript carry Standard Model hypercharge.\label{tab:appendix2}}
\end{table}

As will be specified below, the action of the first orbifold reduces the non-abelian gauge symmetries  
\begin{equation}
SU(6)\times SU(4)/\mathbb{Z}_2 \rightarrow [SU(3)\times SU(2)]^2,
\end{equation}
 at which stage some residual, unwanted fields remain. These are then removed with the second, $\mathbb{\tilde Z}_2$ orbifold, very analoguous to what happens in Scherk-Schwarz supersymmetry breaking. Concretely, following the procedure described in \cite{Craig:2014roa}, we embed the $\mathbb{Z}_2\times \mathbb{\tilde Z}_2$ in the $SU(6)\times SU(4)$ 
\begin{align}
&\mathbb{Z}_2 : \eta \times \gamma_6\otimes \gamma_4\\
&\mathbb{\tilde Z}_2 : \eta \times \gamma_6
\end{align}
with 
\begin{equation}
 \gamma_6=\left(\!\!\begin{array}{cc}\One_3&\\&-\One_3\end{array}\!\!\right)\quad \mathrm{and}\quad \gamma_4=\left(\!\!\begin{array}{cc}\One_2&\\&-\One_2\end{array}\!\!\right)
\end{equation}
and $\eta=+1$ for the $A$ field and $\eta=-1$ for the $B$ fields. After the $\mathbb{Z}_2$ projection, the gauge groups are broken and the only matter fields in table \ref{tab:appendix2} remain. Fields with twin quantum numbers are denoted with a prime as usual. In addition to the usual SM + twin field content, there are two remaining fields in the theory, the $q'_A$ and $q_B$ below the double line in table \ref{tab:appendix2}. These phenomenologically troublesome fields are then removed by the $\mathbb{\tilde Z}_2$ orbifold. One can easily verify that the $\mathbb{\tilde Z}_2$ orbifold does not remove any other fields that were not already projected out by the $\mathbb{Z}_2$ orbifold. We therefore end up with the standard twin Higgs, but with no SM hypercharge for the twin fields.

It is worth noting that although the $g_2$ and $g_3$ gauge couplings are automatically equal in both sectors due to the unified nature of their respective groups, this is not the case for $y_t$ and $g_1$. To enforce this we had to impose a $\mathbb{Z}_2$ exchange symmetry by hand in equation (\ref{apphyperlag}). This is a modest price we must pay with respect to the models in \cite{Craig:2014roa}, in order to gain more flexibility in the hypercharge sector.

\FloatBarrier

\bibliography{vector_twin_bib}

\end{document}